\documentclass[a4paper]{amsart}
\usepackage{epic,eepic,amsmath,amsthm,amssymb}
\usepackage{epsfig,cite,indentfirst,oldgerm}
\usepackage{multicol,boxedminipage}

\begin{document}

\title[Free fermions and plane partitions]
{On free fermions and plane partitions}

\author{O Foda, M Wheeler and M Zuparic}

\address{Department of Mathematics and Statistics,
         University of Melbourne, 
         Parkville, Victoria 3010, Australia.}
\email{foda, mwheeler, mzup@ms.unimelb.edu.au}

\date{}

\newcommand{\field}[1]{\mathbb{#1}}
\newcommand{\C}{\field{C}}
\newcommand{\N}{\field{N}}
\newcommand{\Z}{\field{Z}}

\begin{abstract}
We use free fermion methods to re-derive a result of Okounkov and 
Reshetikhin relating charged fermions to random plane partitions, 
and to extend it to relate neutral fermions to strict plane partitions.
\end{abstract}

\maketitle
\newtheorem{ca}{Figure}
\newtheorem{corollary}{Corollary}
\newtheorem{de}{Definition}
\newtheorem{definition}{Definition}
\newtheorem{example}{Example}
\newtheorem{ex}{Example}
\newtheorem{lemma}{Lemma}
\newtheorem{no}{Notation}
\newtheorem{proposition}{Proposition}
\newtheorem{pr}{Proposition}
\newtheorem{remark}{Remark}
\newtheorem{re}{Remark}
\newtheorem{theorem}{Theorem}
\newtheorem{theo}{Theorem}

\def\ll{\left\lgroup}
\def\rr{\right\rgroup}

\newcommand{\Proof}{\medskip\noindent {\it Proof: }}
\def\no{\nonumber}
\def\ni{\noindent}
\def\proofend{\ensuremath{\square}}
\def\pr{'}
\def\beqa{\begin{eqnarray}}
\def\eeqa{\end{eqnarray}}
\def\ba{\begin{array}}
\def\ea{\end{array}}
\def\gl{\begin{swabfamily}gl\end{swabfamily}}
\def\psis{\psi^{*}}
\def\Psis{\Psi^{*}}
\def\union{\mathop{\bigcup}}
\def\vac{|\mbox{vac}\rangle}
\def\cav{\langle\mbox{vac}|}
\def\dprod{\mathop{\prod{\mkern-29.5mu}{\mathbf\longleftarrow}}}
\def\rprod{\mathop{\prod{\mkern-28.0mu}{\mathbf\longrightarrow}}}
\def\r{\rangle}
\def\l{\langle}
\def\a{\alpha}
\def\b{\beta}
\def\hb{\hat\beta}
\def\d{\delta}
\def\g{\gamma}
\def\e{\epsilon}
\def\tg{\operatorname{tg}}
\def\ctg{\operatorname{ctg}}
 \def\sh{\operatorname{sh}}
 \def\ch{\operatorname{ch}}
\def\cth{\operatorname{cth}}
 \def\th{\operatorname{th}}
\def\eps{\varepsilon}
 \def\la{\lambda}
\def\tla{\tilde{\lambda}}
\def\Gh{\widehat{\Gamma}}
\def\tmu{\tilde{\mu}}
\def\s{\sigma}
\def\sul{\sum\limits}
\def\pl{\prod\limits}
\def\lt({\left(}
\def\rt){\right)}
\def\pd #1{\frac{\partial}{\partial #1}}
\def\const{{\rm const}}
\def\argum{\{\mu_j\},\{\la_k\}} 
\def\umarg{\{\la_k\},\{\mu_j\}} 
\def\prodmu #1{\prod\limits_{j #1 k} \sinh(\mu_k-\mu_j)}
\def\prodla #1{\prod\limits_{j #1 k} \sinh(\lambda_k-\lambda_j)}
\newcommand{\bl}[1]{\makebox[#1em]{}}
\def\tr{\operatorname{tr}}
\def\Res{\operatorname{Res}}
\def\det{\operatorname{det}}

\newcommand{\boldN}{\boldsymbol{N}}
\newcommand{\bra}[1]{\langle\,#1\,|}
\newcommand{\ket}[1]{|\,#1\,\rangle}
\newcommand{\bracket}[1]{\langle\,#1\,\rangle}
\newcommand{\infinity}{\infty}

\renewcommand{\labelenumi}{\S\theenumi.}

\let\up=\uparrow
\let\down=\downarrow
\let\tend=\rightarrow
\hyphenation{boson-ic
             ferm-ion-ic
             para-ferm-ion-ic
             two-dim-ension-al
             two-dim-ension-al
             rep-resent-ative
             par-tition}

\newtheorem{Theorem}{Theorem}[section]
\newtheorem{Corollary}[Theorem]{Corollary}
\newtheorem{Proposition}[Theorem]{Proposition}
\newtheorem{Conjecture}[Theorem]{Conjecture}
\newtheorem{Lemma}[Theorem]{Lemma}
\newtheorem{Example}[Theorem]{Example}
\newtheorem{Note}[Theorem]{Note}
\newtheorem{Definition}[Theorem]{Definition}
                                                                               
\renewcommand{\mod}{\textup{mod}\,}
\newcommand{\wt}{\text{wt}\,}

\newcommand{\T}{{\mathcal T}}
\newcommand{\U}{{\mathcal U}}
\newcommand{\tT}{\tilde{\mathcal T}}
\newcommand{\tU}{\widetilde{\mathcal U}}
\newcommand{\Y}{{\mathcal Y}}
\newcommand{\B}{{\mathcal B}}
\newcommand{\D}{{\mathcal D}}
\newcommand{\M}{{\mathcal M}}
\renewcommand{\P}{{\mathcal P}}
\newcommand{\R}{{\mathcal R}}

\hyphenation{And-rews
             Gor-don
             boson-ic
             ferm-ion-ic
             para-ferm-ion-ic
             two-dim-ension-al
             two-dim-ension-al}

\section{Introduction}\label{introduction}

In \cite{OR}, Okounkov and Reshetikhin observed that certain 
exponentials of bilinears in generators of a Clifford algebra 
generate random plane partitions, and used that observation to 
define and study a new class of stochastic processes. Following 
common usage, as for example in \cite{blue-book}, we refer to 
the generators of a Clifford algebra as {\it free fermions}, 
and to exponentials of bilinears in fermions as {\it vertex 
operators} \footnote{In \cite{OR}, the operators that generate 
plane partitions are referred to as {\it half-vertex operators} 
because each can be interpreted as a specialization of one of 
two factors that together form a {\it fermion vertex operator}. 
Here, we interpret each as a specialization of an {\it evolution 
operator}, and simply refer to them as {\it vertex operators}.}.

Here, we use fermion calculus \footnote{The various algebraic 
methods based on the Clifford algebra of free fermion operators.} 
to take a closer look at the connection between vertex 
operators and plane partitions, and to extend this connection to 
another type of vertex operators and plane partitions. The free 
fermions that appear in \cite{OR} carry a charge \footnote{One can 
think of an electric charge, such as that of an electron, normalized 
to $\pm1$.}, and the corresponding plane partitions are unrestricted. 
The free fermions that appear in the extension discussed in this work 
are neutral, and the corresponding plane partitions are restricted, 
as we will see below. 

\subsection{Contents}
The paper consists of two parts that are written in a way that 
emphasizes their similarities. The first part consists of sections 
{\bf \ref{kp-1}}, 
{\bf \ref{kp-2}} and 
{\bf \ref{kp-3}},
which are devoted to charged fermions (two species of fermions 
are involved) and random plane partitions (with no restrictions). 
In {\bf \ref{kp-1}}, we interpret the operators $\Gamma_{\pm}$ 
used in \cite{OR} as specializations (by setting the variables 
to certain constant values) of evolution operators from 
an integrable hierarchy, based on charged fermions with two 
essential singularities in the spectral 
parameter \cite{jimbo-miwa}, and study their basic properties. 
In {\bf \ref{kp-2}}, we outline a proof, based on fermion calculus, 
that the action of $\Gamma_{\pm}$ on a Young diagram $\mu$, 
generates a Young diagram $\nu$ that {\it interlaces} with $\mu$. 
In \cite{OR}, the interlacing condition was obtained using properties 
of skew Schur functions \cite{macdonald}. The point of a proof based 
on fermion calculus is that it may be amenable to generalizations 
to situations where one based on symmetric functions is not readily 
available.
In {\bf \ref{kp-3}}, we reproduce the result of \cite{OR}, that an 
expectation value of products of $\Gamma_{\pm}$ is MacMahon's 
generating function of random plane partitions \cite{macdonald}, 
and observe that this generating function is a specialization 
of a tau function (a solution of Hirota's bilinear form) of the 
two-dimensional Toda lattice hierarchy. 
We give the details for completeness, and in preparation for deriving 
analogous results in the following sections and in future work. 

The second part consists of sections 
{\bf \ref{bkp-1}}, 
{\bf \ref{bkp-2}} and 
{\bf \ref{bkp-3}}, which are devoted to neutral fermions and a restricted 
class of plane partitions.
In {\bf \ref{bkp-1}}, we recall basic facts related to neutral fermions 
from an integrable hierarchy with two essential 
singularities in the spectral parameter, introduce analogues of the 
evolution operators of {\bf \ref{kp-1}}, and study their basic 
properties.
In {\bf \ref{bkp-2}}, we show that the neutral fermion evolution 
operators, with suitable specializations of the time variables, 
give vertex operators $\Gh_{\pm}$ that act on strict Young diagrams 
(all parts are distinct) to generate interlacing strict Young diagrams. 
In {\bf \ref{bkp-3}}, we use $\Gh_{\pm}$ to generate and count 
a class of plane partitions that satisfy two conditions: 
{\bf A.} Diagonal slices are interlacing strict Young diagrams, 
{\bf B.} Connected horizontal plateaux (which by condition 
{\bf A} are maximally one square wide) are 2-coloured 
\footnote{These results were announced in \cite{fw} and obtained 
independently, using different methods, in \cite{v1} and further 
studied in
\cite{v2}. In this work, we give the details of the fermionic 
approach and comment on the connection to specific integrable 
hierarchies.}. We postulate that the resulting generating function 
is a specialization of a tau function of the neutral fermion 
analogue of the two-dimensional Toda lattice hierarchy. 

\section{Charged fermion vertex operators}\label{kp-1}

\subsection{Charged fermions}

Consider two species of free fermion operators $\{\psi_{m}, 
\psis_{m}\}$, $m \in \Z$, with charges $\{+1, -1\}$ 
(independently of $m$) and energies $m$ (independently of 
species). They generate a Clifford algebra over $\C$ defined
by the anti-commutation relations 
\begin{equation}
\left.
\begin{array}{l}
\left[  \psi_m,  \psi_n \right]_{+} = 0 \\ \\ 
\left[ \psis_m, \psis_n \right]_{+} = 0 \\ \\ 
\left[  \psi_m, \psis_n \right]_{+}= \delta_{m,n}
\end{array}
\right\}\ \forall \ m, n \in \Z
\label{aa}
\end{equation}

We are interested in computing inner products of initial and final 
charged fermion states, and matrix elements of operators that 
interpolate them. A convenient way to perform these computations 
starts from a representation Fock space basis vectors in terms 
of {\it Maya diagrams} \cite{blue-book}.  

\subsection{Maya diagrams}
A Maya diagram is an infinite one-dimensional integral lattice, 
or {\it Go board}.  Each site on the lattice is labeled by its 
position $i \in \Z$. 
A site corresponds to an allowed energy state. The position of 
a site is also the corresponding energy (so in a sense, a Maya 
diagram is a graphical representation of the spectrum of a system). 

On each site we place a {\it black stone} or a {\it white stone}. 
The initial vacuum state is represented by a Maya diagram with 
black stones at the origin ($i=0$), and at all positive 
sites, 
and white stones at all negative sites \footnote{This is the initial 
vacuum state of the sector of the Fock space with net zero charge. 
We will not consider vacuum states with non-zero net charges in this 
work, as they do not lead to different enumerative results.} as 
in Figure {\bf \ref{charged-Maya-vacuum}}.
The final vacuum state is represented by a Maya diagram
with white stones at the origin ($i=0$), and at all negative sites,
and black stones at all positive sites.

%
\begin{center}
\begin{minipage}{4.9in}
\setlength{\unitlength}{0.0008cm}
\renewcommand{\dashlinestretch}{30}
\begin{picture}(4800, 2250)(-2000, 0)
\thicklines
\path(-0600,600)(12600,600)
\path(6000,1200)(06000,000)
\put(00000,600){\whiten\circle{500}}
\put(01200,600){\whiten\circle{500}}
\put(02400,600){\whiten\circle{500}}
\put(03600,600){\whiten\circle{500}}
\put(04800,600){\whiten\circle{500}}
\put(06000,600){\blacken\circle{500}}
\put(07200,600){\blacken\circle{500}}
\put(08400,600){\blacken\circle{500}}
\put(09600,600){\blacken\circle{500}}
\put(10800,600){\blacken\circle{500}}
\put(12000,600){\blacken\circle{500}}
\end{picture}
\begin{ca}
\label{charged-Maya-vacuum}
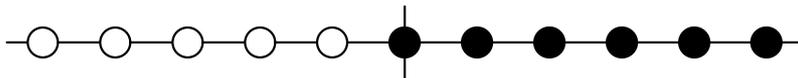
The Maya diagram representation of the initial ground 
state vector in the charged fermion Fock space. The origin is denoted 
by a vertical line.
\end{ca}
\end{minipage}
\end{center}
\bigskip

A finite energy fermion basis vector corresponds to a Maya diagram such 
that, for a sufficiently large $N>0$, all sites located at $i \ge N-1$ 
are occupied by black stones, and sites located at $i \le -N$ are 
occupied by white stones, as in Figure {\bf \ref{charged-Maya-finite}}.

%
\begin{center}
\begin{minipage}{4.9in}
\setlength{\unitlength}{0.0008cm}
\renewcommand{\dashlinestretch}{30}
\begin{picture}(4800, 2000)(-2000, 0)
\thicklines
\path(-0600,0600)(12600,600)
\path(06000,1200)(06000,000)
\put(00000,600){\whiten\circle{500}}
\put(01200,600){\whiten\circle{500}}
\put(02400,600){\blacken\circle{500}}
\put(03600,600){\whiten\circle{500}}
\put(04800,600){\blacken\circle{500}}
%
\put(06000,600){\whiten\circle{500}}
\put(07200,600){\blacken\circle{500}}
\put(08400,600){\blacken\circle{500}}
\put(09600,600){\whiten\circle{500}}
\put(10800,600){\blacken\circle{500}}
\put(12000,600){\blacken\circle{500}}
\end{picture}
\begin{ca}
\label{charged-Maya-finite}
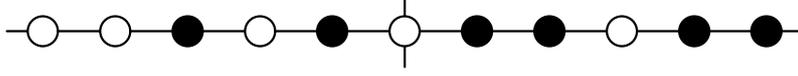
A Maya diagram corresponding to a finite energy 
charged fermion basis vector, where $N \geq 5$. 
\end{ca}
\end{minipage}
\end{center}
\bigskip

\subsection{Charged fermion state vectors} We now translate the
Maya diagrams to the more conventional language of state vectors. 
An initial state vector

\begin{equation}
| j_{1},j_{2},\ldots\rangle, \quad j_{1}<j_{2}<\ldots 
\label{ab}
\end{equation}

\noindent corresponds to a Maya diagram with 
{\it black} stones at sites $\{j_{1},j_{2},\ldots\}$, and white 
stones on all remaining sites. In other words, the corresponding 
Maya diagram is labeled 
by the positions of the {\it black} stones. A final state vector

\begin{equation}
\langle \ldots,i_{2},i_{1}|, \quad \ldots<i_{2}<i_{1} 
\label{ac}
\end{equation}

\noindent corresponds to a Maya diagram with {\it white} stones 
on $\{\ldots,i_{2},i_{1}\}$, and black stones on all remaining 
sites. In other words, the corresponding charged fermion Maya 
diagram is labeled by the positions of the {\it white} stones.

\subsection{Action of charged fermions} $\psi_m$ puts a white stone 
at position $m$ (assuming a black stone is initially there). If there 
is already a white stone at $m$, it annihilates the state 

\begin{eqnarray}
\psi_m | j_1, j_2, \ldots \rangle 
&=& 
\left\{
\begin{array}{ll}
(-)^{k-1} |j_1, \ldots, j_{k-1}, j_{k+1}, \ldots \rangle, &  
m = j_k \bigskip \\ 0, & \mbox{otherwise} 
\end{array}
\right.
\label{ad}
\\
\langle \ldots, i_2, i_1| \psi_m 
&=& 
\left\{
\begin{array}{ll}
(-)^{k} \langle \ldots, i_{k+1}, -m, i_{k}, \ldots, i_1|, & 
i_{k+1}<-m<i_{k} 
\bigskip \\ 0, & \mbox{otherwise} 
\end{array}
\right.
\label{ae}
\end{eqnarray}

Conversely, $\psis_m$ puts a black stone at position $m$ (assuming 
a white stone is initially there). If there is already a black stone 
at $m$, it annihilates the state 

\begin{eqnarray}
\psis_m|j_1,j_2,\ldots\rangle 
&=& 
\left\{
\begin{array}{ll}
(-)^{k}|j_1,\ldots,j_{k},m,j_{k+1},\ldots\rangle, & j_{k}<m<j_{k+1} 
\bigskip  \\ 0, & \mbox{otherwise} 
\end{array}
\right.
\label{af}
\\
\langle \ldots,i_2,i_1|\psis_m 
&=& 
\left\{
\begin{array}{ll}
(-)^{k-1} \langle\ldots,i_{k+1},i_{k-1},\ldots,i_1|, & -m=i_k \bigskip \\ 
0, & \mbox{otherwise} 
\end{array}
\right.
\label{ag}
\end{eqnarray}

The actions in (\ref{ad}--\ref{ag}) respect the anti-commutation 
relations in (\ref{aa}), and define the initial and final vacuum 
states, $|0\rangle$ and $\langle 0|$ by 

\begin{equation}
\left.
\begin{array}{l}
\psi_m |0\rangle = \langle 0| \psi_{n} = 0, \\ \\
\psis_n|0\rangle = \langle 0|\psis_{m} = 0, 
\end{array}
\right\} \quad \forall\ m < 0,\ n \geq 0
\label{ah}
\end{equation}

They also make it possible to create any element of the initial 
or final Fock spaces as a linear combination of

\begin{equation}
\left.
\begin{array}{l}
\phantom{\langle 0|}
\psi_{m_1} \ldots \psi_{m_s} \psis_{n_1} \ldots \psis_{n_r}
|0 \rangle 
\\ \\
\langle 0|
\psi_{n_r}\ldots\psi_{n_1}\psis_{m_s}\ldots\psis_{m_1}
\phantom{|0 \rangle}
\end{array}
\right\} \quad m_1>\ldots>m_s\geq 0, \quad n_1<\ldots<n_r<0
\label{am}
\end{equation}

\noindent respectively. Choosing $\langle 0 | 0 \rangle = 1$, 
we obtain an inner product between the initial and final Fock 
spaces

\begin{equation}
\langle \ldots, i_{2}, i_{1} | j_{1}, j_{2}, \ldots \rangle = 
\prod_{k=1}^{\infty} \delta_{i_k + j_k, 0}
\label{ai}
\end{equation}

\subsection{The Lie algebra $A_\infty$ and charged fermions}
Following \cite{jimbo-miwa}, the algebra $A_\infty$ is the vector 
space 
\begin{equation}
\left\{\sum_{i, j \in \Z} 
a_{ij} :\psi_{i} \psis_j: 
\right\}
\oplus{\C}
\label{aj}
\end{equation}
\noindent equipped with a Lie bracket \footnote{For details, please 
refer to \cite{jimbo-miwa}.}, where the coefficients $a_{ij}$ satisfy 
the condition 
\begin{equation*}
\exists\ N \in \N \ | \ a_{ij}=0, \ \forall \ |i-j|>N  
\end{equation*}
\noindent and the normal-ordered product is defined, as usual, by
\begin{equation*}
:\psi_{i}\psis_{j}:\ = \psi_{i}\psis_{j}-\langle 0|
\psi_{i}\psis_{j}|0\rangle
\end{equation*}

\subsection{$A_{\infty}$ Heisenberg subalgebra} Of particular 
importance are the operators $H_m \in A_\infty$, where  

\begin{equation}
H_m := {\sum_{j \in \Z}:\psi_{j}\psis_{j+m}:}, 
\quad m \in \Z 
\label{ak}
\end{equation}

\noindent which together with the central element $1$ form 
a Heisenberg subalgebra of $A_\infty$

\begin{equation}
\left[ H_m, H_n \right] = m \delta_{m + n, 0}, 
\ \forall \ m, n \in \Z 
\label{al}
\end{equation}

Further, we also have 

\begin{equation}
\left[H_m, \psi_n \right] =  \psi_{-m+n}, \quad\quad 
\left[H_m,\psis_n \right] = -\psis_{m+n}
\label{an}
\end{equation}

Defining the generating functions

\begin{equation}
H_{\pm}(\mathbf{x}) := \sum_{m \in \pm \N} x_m H_m, \quad 
            \Psi(k) := \sum_{j \in \Z} \psi_j k^{j}, \quad 
           \Psis(k) := \sum_{j \in \Z} \psis_j k^{j}
\end{equation}

\noindent and using (\ref{an}), one obtains

\begin{eqnarray}
\left[ H_{\pm}(\mathbf{x}), \Psi(k) \right] 
&=& 
\phantom{-} \sum_{m \in \pm \N} x_m k^{ m} \phantom{{}^{-}} \Psi(k)  
\phantom{{}^{*}} 
:= 
\phantom{-} \xi_{\pm}(\mathbf{x},k) \Psi(k) 
\\ 
\\ 
\left[ H_{\pm}(\mathbf{x}), \Psis(k) \right] 
&=& 
         - \sum_{m \in \pm \N} x_m k^{-m} \Psis(k) 
:= 
-\xi_{\pm}(\mathbf{x}, k^{-1}) \Psis(k)
\label{comm3}
\end{eqnarray}

\subsection{Two charged fermion evolution operators}

The commutators (\ref{comm3}) imply the relations

\begin{eqnarray}
e^{  H_{\pm}(\mathbf{x})}
\Psi(k) 
e^{- H_{\pm}(\mathbf{x})}
&=& 
\Psi(k) 
e^{\xi_{\pm}(\mathbf{x},k)} \label{aq} \\ \nonumber \\ 
e^{  H_{\pm}(\mathbf{x})}\Psis(k) e^{- H_{\pm}(\mathbf{x})}
&=& 
\Psis(k) 
e^{-\xi_{\pm}(\mathbf{x},k^{-1})}  
\nonumber
\end{eqnarray}

This shows that the exponentials $e^{H_{\pm}(\mathbf{x})}$ are 
time evolution operators. 
$H_{+}$ involves the time variables $x_m, m \in \N$, and from 
$\xi_{+}$, we see that the associated essential singularity in 
the spectral parameter $k$ is at $k=\infty$. 
$H_{-}$ involves the time variables $x_{-m}, m \in \N$, and from 
$\xi_{-}$, we see that the associated essential singularity in 
$k$ is at $k=0$. 
This is precisely the situation for integrable hierarchies with 
two essential singularities in the spectral parameter, such as
the two--dimensional Toda lattice hierarchy 
\cite{jimbo-miwa,ueno-takasaki-1}.

\subsection{Specializing the time variables}

So far, the time variables $x_m$ are indeterminates. Setting 

\begin{equation*}
x_m = -\frac{ z^{-m}}{m}, \quad \forall \ m \in \pm \N
\end{equation*}

\noindent where $z$ is an indeterminate, we write 
  $H_{\pm}(\mathbf{x})   := H_{\pm}(z)$, and 
$\xi_{\pm}(\mathbf{x},k) := \xi_{\pm}(z,k)$. 
Then formally

\begin{eqnarray}
\xi_{+}(z,k) = 
- \sum_{m = 1}^{\infty} \frac{1}{m} 
\ll \frac{k}{z} \rr^m &=& 
\phantom{+} \log \ll 1 - \frac{k}{z} \rr \label{ar} \\ 
\nonumber \\ 
\xi_{-}(z,k) =
\phantom{+} 
\sum_{m=1}^{\infty} \frac{1}{m} \ll \frac{z}{k} \rr^m 
&=& 
- \log \ll 1 - \frac{z}{k} \rr 
\nonumber
\end{eqnarray}

\subsection{$A_{\infty}$ vertex operators} Given the above
equation, we define the charged fermion vertex operators, 
$\Gamma_{+}(z)$ and 
$\Gamma_{-}(z)$

\begin{equation}
\Gamma_{+}(z) := \phantom{-} e^{H_{+}(z)} =
\exp{\ll - \sum_{m=1}^{\infty} \frac{z^{-m}}{m} H_m \rr}
\label{ao}
\end{equation} 

\begin{equation}
\Gamma_{-}(z) := e^{-H_{-}(z)} =
\exp{\ll - \sum_{m=1}^{\infty}\frac{z^m}{m}H_{-m} \rr} 
\label{ap}
\end{equation}

Combining these definitions with (\ref{aq}) and 
(\ref{ar}), one finds 

\begin{eqnarray*}
\Gamma_{+}(z) \Psi(k) \phantom{*} \Gamma_{+}^{-1}(z)  & = & 
\ll 1 - \frac{k}{z} \rr  \Psi(k) \\ \\ 
\Gamma_{+}(z) \Psis(k) \Gamma_{+}^{-1}(z) & = & 
\ll \frac{kz}{kz-1} \rr  \Psis(k) \\ \\ 
\Gamma_{-}^{-1}(z) \Psi(k) \phantom{*} \Gamma_{-}(z)  & = & 
\ll \frac{k}{k-z} \rr  \Psi(k) \\ \\
\Gamma_{-}^{-1}(z) \Psis(k) \Gamma_{-}(z) & = & 
\ll 1 - kz \rr \Psis(k)
\label{relations-1}
\end{eqnarray*}

Expanding the relations (\ref{relations-1}) in terms of operators, 
one obtains

\begin{eqnarray*}
\sum_{j \in \Z} \Gamma_{+}(z) \psi_j \Gamma_{+}^{-1}(z) k^{j} & = & 
\sum_{j \in \Z} \psi_j k^{j} \ll 1-\frac{k}{z}\rr \\ \\ 
\sum_{j \in \Z} \Gamma_{+}(z) \psis_j \Gamma_{+}^{-1}(z) k^{j} & = & 
\sum_{j \in \Z} \psis_j k^{j} \ll \ \sum_{n=0}^{\infty} 
\ll \frac{1}{kz} \rr^n \rr \\ \\
\sum_{j \in \Z} \Gamma_{-}^{-1}(z) \psi_j \Gamma_{-}(z) k^{j} & = & 
\sum_{j \in \Z} \psi_j k^{j} \ll \ \sum_{n=0}^{\infty}
\ll \frac{z}{k} \rr^n \rr \\ \\ 
\sum_{j \in \Z} \Gamma_{-}^{-1}(z)\psis_j \Gamma_{-}(z) k^{j} & = & 
\sum_{j \in \Z} \psis_j k^{j} \ll 1 - kz \rr
\end{eqnarray*}

Equating powers of $k$ in the previous expressions gives

\begin{eqnarray*}
\Gamma_{+}(z) \psi_j \Gamma_{+}^{-1}(z) &=& 
\psi_j - \frac{1}{z} \psi_{j-1} 
\\ 
\Gamma_{+}(z) \psis_j \Gamma_{+}^{-1}(z)&=& 
\sum_{n=0}^{\infty} \frac{1}{z^n}\psis_{j+n} 
\\
\Gamma_{-}^{-1}(z)\psi_j \Gamma_{-}(z) &=& 
\sum_{n=0}^{\infty}z^n\psi_{j+n} 
\\ 
\Gamma_{-}^{-1}(z)\psis_j \Gamma_{-}(z)&=& \psis_j-z\psis_{j-1}
\end{eqnarray*}

Given the definitions in (\ref{ao}--\ref{ap}) of 
the vertex operators, one has the commutation relation

\begin{eqnarray*}
\Gamma_{+}(z) \Gamma_{-}(z') &=& e^{H_{+}(z)}e^{-H_{-}(z')}
\\ \\ 
&=& 
e^{[H_{+}(z),-H_{-}(z')]}\ e^{-H_{-}(z')}e^{H_{+}(z)} 
\\ \\ 
&=&e^{[H_{+}(z),-H_{-}(z')]}\ \Gamma_{-}(z') \Gamma_{+}(z)
\end{eqnarray*}

Given that

\begin{eqnarray*}
\left[ H_{+}(z), -H_{-}(z')\right] &=&
\sum_{m = 1}^{\infty}
\sum_{n =1}^{\infty} \frac{1}{mn} z^{-m}{(z')}^{n}
\left[ H_m, H_{-n}\right] 
\\ \\ 
&=& \sum_{m = 1}^{\infty}
    \sum_{n = 1}^{\infty} \frac{1}{m n} z^{-m}(z')^{n} m \delta_{m,n} \\ \\ 
&=& \sum_{m = 1}^{\infty} \frac{1}{m} \ll \frac{z'}{z} \rr^m = 
- \log \ll 1 - \frac{z'}{z} \rr
\end{eqnarray*}

\noindent we find

\begin{equation}
\Gamma_{+}(z) \Gamma_{-}(z') = 
\ll 1-\frac{z'}{z} \rr^{-1}
\Gamma_{-}(z') \Gamma_{+}(z)
\label{aw}
\end{equation}

\noindent which is the basic commutation relation of charged fermion 
vertex operators. 

\section{Young diagrams}\label{kp-2}

Consider the Young diagram of a partition of an integer into integral 
parts as in Figure {\bf \ref{Young-diagram}}. The boxes have 
coordinates $(i, j),\ i, j \geq 1$.

\begin{de} A hook $h$ of a Young diagram is the set of boxes 

\begin{equation*}
h(p, q| j)= 
\left\{
\union_{k=0}^{p}(j + k, j) 
\right\}
\union 
\left\{
\union_{l=0}^{q-1}(j,j+l)
\right\},\quad p \geq 0,\ q \geq 1
\end{equation*}
\end{de}

\noindent and a full Young diagram $\mu$ is the union of 
hooks

\begin{equation}
\mu = \union_{j=1}^{r} h(p_j, q_j| j) 
\label{ax}
\end{equation}

\noindent for some $r \geq 1$, and 
$p_1 > \ldots > p_r \geq 0,\ q_1 > \ldots > q_r \geq 1$.

%
\begin{center}
\begin{minipage}{5.0in}
\setlength{\unitlength}{0.001cm}
\renewcommand{\dashlinestretch}{30}
\begin{picture}(4800, 4200)(-4000, 0)
%
\thicklines
%
\path(0000,0000)(0600,0000)
\path(0000,0600)(1200,0600)
\path(0000,1200)(1200,1200)
\path(0000,1800)(2400,1800)
\path(0000,2400)(2400,2400)
\path(0000,3000)(3000,3000)
\path(0000,3600)(0000,0000)
\path(0000,3600)(3000,3600)
\path(0600,3600)(0600,0000)
\path(1200,3600)(1200,0600)
\path(1800,3600)(1800,1800)
\path(2400,3600)(2400,1800)
\path(3000,3600)(3000,3000)
\end{picture}
\begin{ca}
\label{Young-diagram}
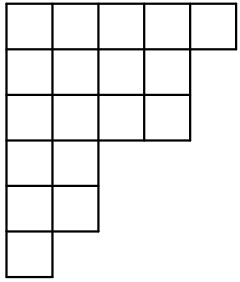
A Young diagram corresponding to the partition of 
$18$ $= 5 +$$ 4 +$$ 4 +$$ 2 +$$ 2 + 1$.
\end{ca}
\end{minipage}
\end{center}
\bigskip

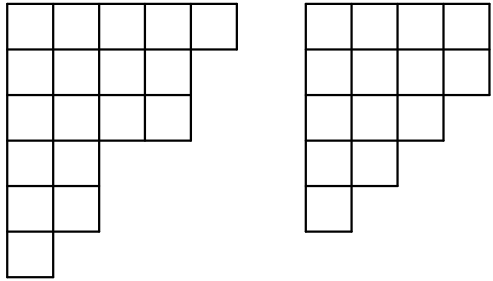
\subsection{Interlacing Young diagrams}

\begin{de} Given the partition $\mu =\{\mu_1,\ldots,\mu_l\}$, 
with $\mu_1 \geq \ldots \geq \mu_l > \mu_{l+1} \equiv 0$, we 
say that the partition $\nu$ interlaces $\mu$, and write $\nu 
\prec \mu$, when 
$\mu_j \geq \nu_j$$\geq \mu_{j+1}$, $\forall\ 1 \leq j $$\leq l$. 
\end{de}

An example of interlacing Young diagrams is in Figure 
{\bf \ref{interlacing}}.

%
\begin{center}
\begin{minipage}{4.0in}
\setlength{\unitlength}{0.001cm}
\renewcommand{\dashlinestretch}{30}
\begin{picture}(4800, 4500)(-2000, 0)
\thicklines
\path(0000,0000)(0600,0000)
\path(0000,0600)(1200,0600)
\path(0000,1200)(1200,1200)
\path(0000,1800)(2400,1800)
\path(0000,2400)(2400,2400)
\path(0000,3000)(3000,3000)
\path(0000,3600)(0000,0000)
\path(0600,3600)(0600,0000)
\path(1200,3600)(1200,0600)
\path(0000,3600)(3000,3600)
\path(1800,3600)(1800,1800)
\path(2400,3600)(2400,1800)
\path(3000,3600)(3000,3000)
\path(3900,0600)(4500,0600)
\path(3900,1200)(5100,1200)
\path(3900,1800)(5700,1800)
\path(3900,2400)(6300,2400)
\path(3900,3000)(6300,3000)
\path(3900,3600)(3900,0600)
\path(3900,3600)(6300,3600)
\path(4500,3600)(4500,0600)
\path(5100,3600)(5100,1200)
\path(5700,3600)(5700,1800)
\path(6300,3600)(6300,2400)
\end{picture}
\begin{ca}
\label{interlacing}
Interlacing Young diagrams.
\end{ca}
\end{minipage}
\end{center}
\bigskip

Let $\mu$ be the partition given by (\ref{ax}), and 
consider the set of diagrams

\begin{equation*}
\mathcal{D}_\mu := 
\left\{\union_{j=1}^{r} h({p'}_j, {q'}_j|j) 
\left|
\right.  
p_j \geq {p'}_j \geq p_j-1,\ q_j \geq 
{q'}_j \geq q_{j+1}+1, \forall \ 1 \leq j \leq r \right\}
\end{equation*}

\noindent where $q_{r+1} \equiv 0$ and $h(-1, {q'}_r|r) 
\equiv \emptyset$. 
Furthermore, let $\mathcal{Y}_\mu \subseteq \mathcal{D}_\mu$ be the 
subset of all diagrams in $\mathcal{D}_\mu$ which are Young diagrams. 
Then $\mathcal{Y}_\mu$ is exactly the set of all Young diagrams that 
interlace $\mu$. In other words

\begin{equation*}
\mathcal{Y}_\mu = \left\{ \nu \right| \left. \nu \prec \mu \right\}
\end{equation*}

\subsection{Generating interlacing Young diagrams} 
Every Maya diagram in (the charge-0 sector of) the initial Fock space 
may be represented uniquely in the form 

\begin{equation*}
|\mu \r :=
(-)^{\kappa} 
\psi_{m_1} \ldots \psi_{m_r} \psis_{n_{1}} \ldots \psis_{n_{r}} |0\r 
= 
(-)^{\kappa}
\dprod_{j=1}^{r}\psi_{m_j} \dprod_{k=1}^{r}\psis_{n_k}
|0\r
\end{equation*}

\noindent and every Maya diagram in (the charge-0 sector of) the final 
Fock space may be represented uniquely in the form

\begin{equation*}
\l \mu | :=
(-)^{\kappa}
\l 0| 
\psi_{n_r} \ldots 
\psi_{n_1} \psis_{m_r} \ldots 
\psis_{m_1} 
= 
(-)^{\kappa}\l 0|
\rprod_{j=1}^{r}\psi_{n_j}
\rprod_{k=1}^{r}\psis_{m_k}
\end{equation*}

\noindent $m_1>\ldots>m_r \geq 0,\ n_1 <\ldots<n_r<0$, 
with $\kappa := \sum_{k=1}^{r}(m_k+k)$. The arrows on the products on 
the right hand sides indicate that the fermion operators in the product 
are ordered as shown explicitly on the left hand sides (which makes 
a difference as these are anti-commuting operators). 

Therefore, every (charge-0) Maya diagram in the initial (respectively, 
final) Fock space is uniquely associated with a set of integers 
$ m_1 > \ldots > m_r \geq 0,\ n_1 < \ldots < n_r < 0$, and by 
choosing the integers in (\ref{ax}) to be 

\begin{equation*}
p_j=m_j, \quad q_j=-n_j, \quad \forall\ 1 \leq j \leq r
\end{equation*}

\noindent we obtain a bijection between any (charge-0) Maya diagram 
in the initial (respectively, final) Fock space and the corresponding 
partition.

\begin{lemma} Let $| \mu \r$ and $\l \mu |$ be the initial and 
final state vectors corresponding to the partition $\mu$. Then 

\begin{equation}
\langle\nu| \Gamma_{+}(z)| \mu \rangle = \left\{
\begin{array}{ll}
z^{|\nu|-|\mu|}, & \quad\nu \prec \mu \bigskip \\ 
0, & \quad \mbox{otherwise}
\end{array}
\right. 
\label{bb}
\end{equation}

\begin{equation}
\langle\mu| \Gamma_{-}(z)| \nu \rangle = 
\left\{
\begin{array}{ll}
z^{|\mu|-|\nu|}, & \quad\nu \prec \mu \bigskip \\ 
0, & \quad \mbox{otherwise}
\end{array}
\right.
\label{bc}
\end{equation}
\end{lemma}

\noindent{\bf Proof.} 

\begin{eqnarray}
\Gamma_{+}(z)|\mu\r &=& (-)^{\kappa} \Gamma_{+}(z)
\dprod_{j=1}^{r}\psi_{m_j} 
\dprod_{k=1}^{r}\psis_{n_k}|0\r 
\no \\ 
\no \\ 
&=& (-)^{\kappa}
\dprod_{j=1}^{r}
\ll
\Gamma_{+}(z)\psi_{m_j}\Gamma_{+}^{-1}(z)
\rr
\dprod_{k=1}^{r}
\ll
\Gamma_{+}(z)\psis_{n_k} \Gamma_{+}^{-1}(z)
\rr
| 0 \r 
\no \\ 
\no \\ 
&=& (-)^{\kappa}
\dprod_{j=1}^{r}
\ll
\psi_{m_j}-\frac{1}{z}\psi_{(m_j-1)}
\rr
\dprod_{k=1}^{r}
\ll
\sum_{i=0}^{\infty}\frac{1}{z^i}
\psis_{(n_k+i)}
\rr
|0\r 
\label{bd} 
\end{eqnarray}
Consider the action of the second product from the left in the 
above equation, which we call $P$ on the vacuum $|0\r$.
\begin{eqnarray*}
P |0\r
&=&
\dprod_{k=1}^{r}
\ll
\sum_{i=0}^{\infty}\frac{1}{z^i}
\psis_{(n_k+i)}
\rr
|0\r 
\\
&=&
\ll 
\sum_{i=0}^{\infty}\frac{1}{z^i}
\psis_{(n_1+i)}
\rr
\ll 
\sum_{i=0}^{\infty}\frac{1}{z^i}
\psis_{(n_{2}+i)}
\rr
\ldots
\ll 
\sum_{i=0}^{\infty}\frac{1}{z^i}
\psis_{(n_r+i)}
\rr
|0\r
\no
\end{eqnarray*}
Split the first sum from the left into two parts to obtain
\begin{eqnarray*}
P |0\r
=
\ll
\sum_{i=0}^{-n_1+n_2-1}
\frac{1}{z^i}
\psis_{(n_1+i)}
+
\frac{1}{z^{n_2-n_1}}
\sum_{i=0}^{\infty}
\frac{1}{z^i}
\psis_{(n_2+i)}
\rr
\\
\times
\ll 
\sum_{i=0}^{\infty}\frac{1}{z^i}
\psis_{(n_{2}+i)}
\rr
\ldots
\ll 
\sum_{i=0}^{\infty}\frac{1}{z^i}
\psis_{(n_r+i)}
\rr
|0\r
\no
\end{eqnarray*}
Using the identity
\begin{equation*}
\ll 
\sum_{i=0}^{\infty}\frac{1}{z^i}
\psis_{(n+i)}
\rr
\ll 
\sum_{i=0}^{\infty}\frac{1}{z^i}
\psis_{(n+i)}
\rr
=
0
\end{equation*}
which follows directly by expanding the sums and using 
the anti-commutation relation (\ref{aa}), we obtain
\begin{eqnarray*}
P |0\r
=
\ll
\sum_{i=0}^{-n_1+n_2-1}
\frac{1}{z^i}
\psis_{(n_1+i)}
\rr
\ll 
\sum_{i=0}^{\infty}\frac{1}{z^i}
\psis_{(n_{2}+i)}
\rr
\ldots
\ll 
\sum_{i=0}^{\infty}\frac{1}{z^i}
\psis_{(n_r+i)}
\rr
|0\r
\no
\end{eqnarray*}
This procedure can then be performed on the second sum from the
left, and so on, until one reaches the last sum, which truncates
using the fact that $\psis_{n}$ annihilates the vacuum $|0\r$ 
for all $n \geq 0$. Hence
\begin{eqnarray*}
P |0\r
=
\ll 
\sum_{i=0}^{-n_1+n_2-1}\frac{1}{z^i}
\psis_{(n_1+i)}
\rr
\ll 
\sum_{i=0}^{-n_{2}+n_3-1}\frac{1}{z^i}
\psis_{(n_{2}+i)}
\rr
\ldots
\ll 
\sum_{i=0}^{-n_r-1}\frac{1}{z^i}
\psis_{(n_r+i)}
\rr
|0\r
\no
\end{eqnarray*}
Using the above result in (\ref{bd}), we obtain
\begin{equation}
\Gamma_{+}(z)|\mu\r
=
(-)^{\kappa}
\dprod_{j=1}^{r}
\ll
\psi_{m_j}-\frac{1}{z}\psi_{(m_j-1)}
\rr
\dprod_{k=1}^{r}
\ll
\sum_{i=0}^{-n_k+n_{(k+1)}-1}\frac{1}{z^i}\psis_{(n_k+i)}
\rr |0\r
\label{result1}
\end{equation}

\noindent
where we have defined $n_{r+1} = 0$. Analogously to the proof of 
(\ref{result1}), we have 

\begin{eqnarray}
\langle\mu|
\Gamma_{-}(z) &=& (-)^{\kappa}\l 0|
\rprod_{j=1}^{r}\psi_{n_j}
\rprod_{k=1}^{r}\psis_{m_k}
\Gamma_{-}(z) \no \\ \no \\ &=& (-)^{\kappa}\l 0|
\rprod_{j=1}^{r}
\ll
\Gamma_{-}^{-1}(z)\psi_{n_j}
\Gamma_{-}(z)
\rr
\rprod_{k=1}^{r}
\ll
\Gamma_{-}^{-1}(z)\psis_{m_k}
\Gamma_{-}(z)
\rr \no \\ \no \\ &=& (-)^{\kappa}\l 0|
\rprod_{j=1}^{r}
\ll
\sum_{i=0}^{\infty}z^i \psi_{(n_j+i)}
\rr
\rprod_{k=1}^{r}
\ll
\psis_{m_k}-z\psis_{(m_k-1)}
\rr
\no \\ \no \\ &=& (-)^{\kappa}\l 0|
\rprod_{j=1}^{r}
\ll
\sum_{i=0}^{-n_j+n_{j+1}-1}z^i \psi_{(n_j+i)}
\rr
\rprod_{k=1}^{r}
\ll
\psis_{m_k}-z \psis_{(m_k-1)}
\rr
\label{be}
\end{eqnarray}

It is readily seen that exactly all of the Young diagrams present 
in $\mathcal{Y}_\mu$ are reproduced in the expansions of 
(\ref{result1}) and (\ref{be}), but with weighting factors of $z$. 
Letting

\begin{eqnarray*}
m_j \geq {m'}_j \geq m_j-1,
\quad -n_j \geq -{n'}_j \geq -n_{j+1}+1,\quad 
\forall\ 1 \leq j \leq r \\ \\ 
n_{r+1}\equiv 0, \quad h(-1,-{n'}_r|r) \equiv \emptyset 
\quad\quad\quad\quad\quad\quad
\end{eqnarray*}

\noindent (\ref{bd}) is a sum of all weighted Young diagrams 
of the form

\begin{equation*}
\prod_{j=1}^{r}z^{{m'}_j-m_j} z^{-{n'}_j+n_j}
\union_{k=1}^{r}h({m'}_k, -{n'}_k| k)
\end{equation*}

\noindent whilst (\ref{be}) is a sum of all weighted 
Young diagrams of the form

\begin{equation*}
\prod_{j=1}^{r}z^{m_j-{m'}_j}z^{-n_j+{n'}_j}
\union_{k=1}^{r}h({m'}_k,-{n'}_k|k)
\end{equation*}

In other words

\begin{eqnarray*}
\Gamma_{+}(z)|\mu\r &=& \sum_{\nu \prec \mu} z^{|\nu|-|\mu|}|\nu\r \\ \\ 
\l\mu |\Gamma_{-}(z) &=& \sum_{\nu\prec\mu}z^{|\mu|-|\nu|}\l\nu | 
\quad\quad\proofend
\end{eqnarray*}

The above results are known, but in \cite{OR} they were obtained 
using the relationship of charged vertex operators and skew Schur 
functions and the properties of the latter. The above proofs rely 
only on fermion calculus.

\section{Plane partitions}\label{kp-3}

In this section, following \cite{OR}, 
we use the above result, namely that charged fermion vertex 
operators act on Young diagrams to produce new Young diagrams 
that interlace with the first, to count plane partitions. An 
example of a plane partition is in Figure 
{\bf \ref{plane-partition}}.
The basic observation here, also due to 
\cite{OR}, is that adjacent diagonal slices of plane partitions  
are interlacing Young diagrams. The diagonal slices of the plane
partition in Figure {\bf \ref{plane-partition}} are shown on 
a planar representation of the same plane partition in Figure 
{\bf \ref{slices}}.

\begin{de} A plane partition $\pi$ is a collection of integers 
$\pi(i,j)\geq 0$ assigned to each of the coordinate-labelled 
boxes $(i,j)$, restricted by the conditions 
$\pi(i+1,j) \leq \pi(i,j)$, 
$\pi(i,j+1) \leq \pi(i,j)$, 
$\forall\ i,j \geq 1$,
and the finiteness condition
$\lim_{i \rightarrow \infty} \pi(i,j)$ 
$=$ 
$\lim_{j \rightarrow \infty} \pi(i,j)$ 
$= 0$.
\end{de}

%
\begin{center}
\begin{minipage}{2.5in}
\setlength{\unitlength}{0.001cm}
\renewcommand{\dashlinestretch}{30}
\begin{picture}(4800, 5000)(0, 0)
\thicklines
\path(3900,1500)(4500,1500)
\path(1800,4200)(2400,4200)
\path(1800,4200)(1500,3900)
\path(2400,4200)(2100,3900)
\path(1500,3900)(2100,3900)
\path(1500,3900)(1500,3300)
\path(2100,3900)(2100,3300)
\path(1500,3300)(2100,3312)
\path(2400,4200)(2400,3600)
\path(2400,3600)(2100,3300)
\path(1500,3300)(1200,3000)
\path(2100,3300)(1800,3000)
\path(1200,3000)(1800,3000)
\path(1200,3000)(1200,2400)
\path(1800,3000)(1800,2400)
\path(1200,2400)(1800,2400)
\path(1200,2400)(0900,2100)
\path(1800,2400)(1500,2100)
\path(0900,2100)(1500,2100)
\path(0900,2100)(0900,1500)
\path(1500,2100)(1500,1500)
\path(0900,1500)(1500,1500)
\path(0900,1500)(0600,1200)
\path(1500,1500)(1200,1200)
\path(0600,1200)(1200,1200)
\path(0600,1200)(0600,0600)
\path(1200,1200)(1200,0600)
\path(0600,0600)(1200,0600)
\path(0600,0600)(0000,0000)
\path(2100,3300)(2100,2700)
\path(2100,2700)(1800,2400)
\path(1800,2400)(1800,1800)
\path(1800,1800)(1500,1500)
\path(1500,1500)(1500,0900)
\path(1500,0900)(1200,0600)
\path(2400,3600)(2400,3000)
\path(2400,3000)(2100,2700)
\path(2400,3000)(3000,3000)
\path(2100,2700)(2700,2700)
\path(3000,3000)(2700,2700)
\path(1800,2400)(2400,2400)
\path(2700,2700)(2400,2400)
\path(2400,2400)(2400,1800)
\path(1800,1800)(2400,1800)
\path(2400,1800)(2100,1500)
\path(1500,1500)(2100,1500)
\path(2100,1500)(2100,0900)
\path(1500,0900)(2100,0900)
\path(3000,3000)(3000,2400)
\path(2700,2700)(2700,2100)
\path(3000,2400)(2400,1800)
\path(3000,2400)(4500,2400)
\path(3600,2400)(2700,1500)
\path(2700,2100)(3300,2100)
\path(2400,1800)(3000,1812)
\path(2100,1500)(2700,1500)
\path(2700,1500)(2700,0900)
\path(2100,0900)(2700,0900)
\path(4200,2400)(3900,2100)
\path(4500,2400)(4800,2400)
\path(4800,2400)(4500,2100)
\path(3300,2100)(4500,2100)
\path(3000,1800)(3000,1200)
\path(3900,2100)(3900,1500)
\path(4500,2100)(4500,1500)
\path(4800,2400)(4800,1800)
\path(4800,1800)(4500,1500)
\path(4800,1800)(5400,1800)
\path(3900,2100)(3600,1800)
\path(3000,1800)(3600,1800)
\path(3600,1800)(3600,1200)
\path(3000,1200)(3600,1200)
\path(3900,1500)(3600,1200)
\path(3000,1200)(2700,0900)
\path(1800,4800)(1800,4200)
\end{picture}
\begin{ca}
\label{plane-partition}
A plane partition. 
\end{ca}
\end{minipage}
\end{center}
\bigskip

\subsection{Diagonal slices}

The interlacing condition on Young diagrams can be used to construct 
plane partitions. Given a plane partition $\pi$, decompose it 
into its diagonal slices, which are the Young diagrams $\mu_m$, $m 
\in \Z$:

\begin{equation*}
\mu_m = \left\{
\begin{array}{ll}
\{\pi(-m+1,1),\ \pi(-m+2,2),\ldots\}, & \quad 
m \leq 0 \\ \\
\{\pi(1,m+1),\ \pi(2,m+2),\ldots\}, & \quad 
m \geq 0
\end{array}
\right.
\end{equation*}

%
\begin{center}
\begin{minipage}{4.9in}
\setlength{\unitlength}{0.0012cm}
\renewcommand{\dashlinestretch}{30}
\begin{picture}(4800, 4000)(-3000, 0)
\thicklines
\path(0000,0600)(0600,0600)
\path(0000,1200)(1800,1200)
\path(0000,1800)(2400,1800)
\path(0000,2400)(3000,2400)
\path(0000,3000)(0000,0600)
\path(0000,3000)(3000,3000)
\path(0600,3000)(0600,0600)
\path(1200,3000)(1200,1200)
\path(1800,3000)(1800,1200)
\path(2400,3000)(2400,1800)
\path(3000,3000)(3000,2400)
\thinlines
\path(0000,1200)(0900,0300)
\path(0000,1800)(0900,0900)
\path(0000,2400)(1500,0900)
\path(0000,3000)(2100,0900)
\path(0600,3000)(2100,1500)
\path(1200,3000)(2700,1500)
\path(1800,3000)(2700,2100)
\path(2400,3000)(3300,2100)
\put(0300,0900){1}
\put(0300,1500){2}
\put(0300,2100){3}
\put(0300,2700){4}
\put(0900,1500){1}
\put(0900,2100){2}
\put(0900,2700){2}
\put(1500,1500){1}
\put(1500,2100){1}
\put(1500,2700){1}
\put(2100,2100){1}
\put(2100,2700){1}
\put(2700,2700){1}
\put(0900,0600){$\mu_{-2}$}
\put(0900,0000){$\mu_{-3}$}
\put(1500,0600){$\mu_{-1}$}
\put(2100,1200){$\mu_{ 1}$}
\put(2100,0600){$\mu_{ 0}$}
\put(2700,1800){$\mu_{ 3}$}
\put(2700,1200){$\mu_{ 2}$}
\put(3300,1800){$\mu_{ 4}$}
\end{picture}
\begin{ca}
\label{slices}
A planar representation of the plane partition in Figure 
{\bf \ref{plane-partition}} and its diagonal slices. 
The integers in the boxes are the heights of the corresponding
columns.
\end{ca}
\end{minipage}
\end{center}
\bigskip
Then successive diagonal slices of $\pi$ are interlacing 
Young diagrams, in the sense that

\begin{equation}
\emptyset = \mu_{-M} \prec \ldots \prec \mu_{-1} \prec \mu_0 \succ \mu_1 
\succ 
\ldots 
\succ 
\mu_{N} = \emptyset
\label{bf}
\end{equation}

\noindent for sufficiently large $M,N \in \N$. Since we understand 
interlacing Young diagrams in the context of fermion calculus, we can 
use (\ref{bf}) to construct plane partitions in the same 
formalism. Consider, for instance, the scalar product 

\begin{eqnarray}
S_{A}(q) &:=& \langle 0|{\rprod_{j=1}^{\infty}
\Gamma_{+}\ll q^{\frac{-2j+1}{2}}\rr}\ 
{\dprod_{k=1}^{\infty}
\Gamma_{-}\ll q^{\frac{2k-1}{2}}\rr}|0\rangle \label{bh} 
\\ 
\no \\ 
&=&
\sum_{\mu}\langle 0|
{\rprod_{j=1}^{\infty}
\Gamma_{+}
\ll q^{\frac{-2j+1}{2}}\rr}|\mu\rangle\langle\mu|\ 
{\dprod_{k=1}^{\infty}
\Gamma_{-}\ll q^{\frac{2k-1}{2}}\rr}|0
\rangle 
\label{bg} 
\end{eqnarray}

\noindent where $q$ is an indeterminate and $\sum_{\mu}$ denotes 
a sum over all Young diagrams $\mu$. We know that 
$\Gamma_{+}(z)|\mu\rangle$ and $\langle\mu|\Gamma_{-}(z)$ generate 
all Young diagrams $\nu$ that are interlacing with $\mu$, with 
weightings given by (\ref{bb}--\ref{bc}). It follows 
that (\ref{bg}) generates a weighting equal to

\begin{equation*}
{\rprod_{j=1}^{M}\langle\nu_{-j}|
\Gamma_{+} \ll q^{\frac{-2j+1}{2}}\rr}|\nu_{-j+1}\rangle\ 
{\dprod_{k=1}^{N}\langle\nu_{k-1}|
\Gamma_{-}\ll q^{\frac{2k-1}{2}}\rr}|\nu_k\rangle = 
\prod_{j=-M}^{N}q^{|\nu_j|}
\end{equation*}

\noindent for all sequences of Young diagrams of the form 

\begin{equation*}
\emptyset=\nu_{-M} \prec \ldots 
\prec \nu_{-1} \prec \nu_0 
\succ \nu_1 \ldots \succ \nu_N=\emptyset
\end{equation*}

\noindent which, as we know, are in one-to-one correspondence with plane 
partitions. Hence 

\begin{equation*}
S_{A}(q)=\sum_{\pi}q^{|\pi|}
\end{equation*}  

\noindent where $\sum_\pi$ denotes a sum over all plane partitions $\pi$, 
and $|\pi|$ is the weight of the plane partition (the number of boxes). 
Applying the commutation relation (\ref{aw}) successively to (\ref{bh}), 
one recovers MacMahon's product form of the generating function $S_{A}(q)$:

\begin{equation}
S_{A}(q) = \prod_{n=1}^{\infty}
\ll
\frac{1}{1-q^n}
\rr^n
\label{sa-q}
\end{equation}

This is the result of \cite{OR}.
This ends our review of known results in the context of integrable 
hierarchies based on charged fermions, and their derivation in the 
language of fermion calculus.

\subsection{The charged fermion two-dimensional Toda lattice hierarchy}

Let ${\bf z} = \{z_1, z_2, \cdots\}$ be an infinite set of variables, 
and $s_{{}_Y}({\bf z})$ be the Schur function associated to the Young 
diagram $Y$ \cite{macdonald}. The generating function in (\ref{sa-q}) 
is the specialization of 

\begin{equation}
S_A = \sum_{Y} 
s_{Y} (x_1, x_2, \ldots)
s_{Y} (y_1, y_2, \ldots)
=
\prod_{i, j=1}^{\infty} 
\frac{1}{1 - x_i y_j} 
\label{sa}
\end{equation}

\noindent obtained by setting 
$\{{\bf x}\}$ 
$=$ 
$\{{\bf y}\}$ 
$=$ 
$\{q^{\frac{1}{2}}, q^{\frac{3}{2}}, \ldots \}$. 
Under the change of variables 

$$
x_n' = \sum_{j=1}^{\infty} \frac{x^n_j}{n},
\quad
y_n' = \sum_{j=1}^{\infty} \frac{y^n_j}{n}
$$ 

\ni one can rewrite (\ref{sa}) as

\begin{equation}
S_A = 
\sum_{Y} 
{\chi}_{{}_Y} (x_1', x_2', \ldots)
{\chi}_{{}_Y} (y_1', y_2', \ldots)
\label{sa2}
\end{equation}

\ni where $\chi_{{}_{Y}}$ is the character polynomial associated 
to $Y$ \cite{macdonald}. $S_A$ is related to the two-dimensional 
Toda hierarchy as follows. From \cite{takasaki-1}, the general 
solution of the initial value problem of the two-dimensional Toda 
lattice is a tau function of the form 

\begin{equation}
\tau(s, {\bf x'}, {\bf y'}) = 
\sum_{Y_1, Y_2 \subset (s-m)\times(n-s)} 
A(s, Y_1, Y_2) 
\chi_{{}_{Y_1}} ({x_1'}, {x_2'}, \ldots)
\chi_{{}_{Y_2}} ({y_1'}, {y_2'}, \ldots)
\label{toda}
\end{equation}
\noindent where $s \in \Z$ labels the lattice sites, $Y_1$ and $Y_2$ 
are Young diagrams restricted to a rectangle of size 
$(s-m)\times(n-s)$, and $A(s, Y_1, Y_2)$ are scalar coefficients 
that encode the initial value data as defined explicitly in 
\cite{takasaki-1}. It is straightforward to start from (\ref{toda}), 
set $A(s, Y_1, Y_2) = \delta_{Y_1, Y_2}$, take the limits 
$m \rightarrow - \infty$ and $n \rightarrow \infty$, while 
satisfying the two-dimensional Toda bilinear identity, and 
show that (\ref{toda}) is a tau function \footnote{This connection 
to the two-dimensional Toda lattice hierarchy is subtle because 
taking the large lattice limit is non-trivial \cite{takasaki-1}.}.

\section{Neutral fermion vertex operators}\label{bkp-1}

\subsection{Neutral fermions} 

In this section, we recall analogues of the properties discussed 
in {\bf \ref{kp-1}}, but now for neutral fermions. 
Following \cite{jimbo-miwa}, we define the neutral fermion 
operators $\{\phi_m\}$ in terms of $\{\psi_m, \psis_m\}$  
\begin{equation}
\phi_m := \frac{1}{\sqrt{2}} \ll \psi_m + (-)^m \psis_{-m} \rr, 
\quad
m \in \Z
\label{phi}
\end{equation}

\noindent which are linear combinations of the charged fermions that 
remain invariant under the isomorphisms that define the generators 
of the algebra $B_\infty$ as a subalgebra of $A_\infty$. More precisely, 
$B_\infty$ is generated by all $X \in A_{\infty}$ such that 
$\sigma_{0}(X) = X$, where
$\sigma_{0}(\psi_m)  := (-)^{m} \psis_{-m}$ and 
$\sigma_{0}(\psis_m) := (-)^{m}  \psi_{-m}$.

Given (\ref{phi}), it is easy to show that the neutral fermions 
satisfy the anti-commutation relation

\begin{equation}
\left[ \phi_m, \phi_n \right]_{+} = (-)^m \delta_{m + n, 0} 
\label{neutralanti}
\end{equation}

\subsection{Half-line Maya diagrams} Considering the neutral 
fermions in their own right, one can show that only one half 
of a Maya diagram is modified under the action of $\phi_m$. 
In other words, for neutral fermions, we can use the usual 
Maya diagrams of {\bf \ref{kp-1}}, but with the following 
restrictions.

All initial states are such that there are {\it white stones} 
at all negative sites. In the initial vacuum state there are 
black stones at the origin and all positive sites, as in Figure 
{\bf \ref{neutral-Maya-vacuum}}.

%
\begin{center}
\begin{minipage}{4.9in}
\setlength{\unitlength}{0.0008cm}
\renewcommand{\dashlinestretch}{30}
\begin{picture}(4800, 1800)(-2000, 0)
\thicklines
\path(-0600,0600)(12600,600)
\path( 6000,1200)(06000,000)
\put(00000,600){\whiten\circle{500}}
\put(01200,600){\whiten\circle{500}}
\put(02400,600){\whiten\circle{500}}
\put(03600,600){\whiten\circle{500}}
\put(04800,600){\whiten\circle{500}}
\thinlines
\path(-0400,0200)(00400,1000)
\path( 0800,0200)(01600,1000)
\path( 2000,0200)(02800,1000)
\path( 3200,0200)(04000,1000)
\path( 4400,0200)(05200,1000)
\put(06000,600){\blacken\circle{500}}
\put(07200,600){\blacken\circle{500}}
\put(08400,600){\blacken\circle{500}}
\put(09600,600){\blacken\circle{500}}
\put(10800,600){\blacken\circle{500}}
\put(12000,600){\blacken\circle{500}}
\end{picture}
\begin{ca}
\label{neutral-Maya-vacuum}
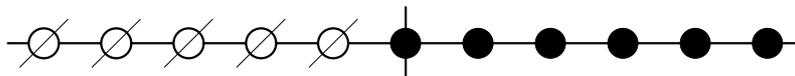
The Maya diagram representation of the ground state vector in the 
neutral fermion Fock space. The site at the origin is denoted with 
a vertical line. All sites to the left of the origin are frozen to 
be white. This is indicated with the diagonal lines.
\end{ca}
\end{minipage}
\end{center}
\bigskip

Finite energy initial states contain a finite number of white 
stones at the origin and/or finite distance positive sites. 
An example is in Figure {\bf \ref{neutral-Maya-finite}}.

%
\begin{center}
\begin{minipage}{4.9in}
\setlength{\unitlength}{0.0008cm}
\renewcommand{\dashlinestretch}{30}
\begin{picture}(4800, 1800)(-2000, 0)
\thicklines
\path(-0600,600)(12600,600)
\path(6000,1200)(6000,000)
\put(00000,600){\whiten\circle{500}}
\put(01200,600){\whiten\circle{500}}
\put(02400,600){\whiten\circle{500}}
\put(03600,600){\whiten\circle{500}}
\put(04800,600){\whiten\circle{500}}
%
\put(06000,600){\whiten\circle{500}}
\put(07200,600){\blacken\circle{500}}
\put(08400,600){\blacken\circle{500}}
\put(09600,600){\whiten\circle{500}}
\put(10800,600){\blacken\circle{500}}
\put(12000,600){\blacken\circle{500}}
\thinlines
\path(-0400,0200)(00400,1000)
\path( 0800,0200)(01600,1000)
\path( 2000,0200)(02800,1000)
\path( 3200,0200)(04000,1000)
\path( 4400,0200)(05200,1000)
\end{picture}
\begin{ca}
\label{neutral-Maya-finite}
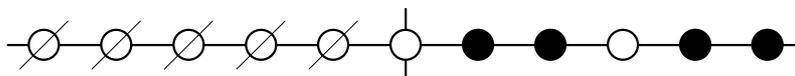
A Maya diagram corresponding to a finite energy neutral fermion basis 
vector. All sites to the left of the origin are frozen to be white. 
\end{ca}
\end{minipage}
\end{center}
\bigskip
Accordingly, all final states are such that there are {\it black stones} 
at all positive sites. In the final vacuum state there are white stones 
at the origin and all negative sites. Finite energy final states contain 
a finite number of black stones at the origin and/or finite distance 
negative sites. We will refer to the Maya diagrams that are relevant to 
neutral fermion states as {\it half-line Maya diagrams}.

\subsection{Neutral fermion state vectors}

Given the above definition, the state vectors $(\ref{ab})-(\ref{ac})$ in 
{\bf \ref{kp-1}} remain as before, but now we have

\begin{equation*}
0 \leq j_1 < j_2 < \ldots, \quad  
\ldots < i_2 < i_1 \leq 0
\end{equation*} 

\subsection{Action of neutral fermions} 

For $m > 0$, $\phi_{m}$ puts a white stone at 
position $m$ (assuming a black stone is initially there), 
otherwise it annihilates the state 

\begin{equation}
\phi_{(m>0)}|j_1,j_2,\ldots\rangle = \left\{
\begin{array}{ll}
(-)^{m+k-1}| j_1, \ldots, j_{k-1}, j_{k+1}, \ldots \rangle, &  
m=j_k \bigskip \\ 0, & \mbox{otherwise} 
\end{array}
\right.
\label{bk}
\end{equation}
and
\begin{equation}
\langle \ldots,i_2,i_1|\phi_{(m>0)} = \left\{
\begin{array}{ll}
(-)^{m+k} \langle\ldots,i_{k+1},-m,i_{k},\ldots,i_1|, & i_{k+1}<-m<i_{k} 
\bigskip \\ 0, & \mbox{otherwise} 
\end{array}
\right.
\label{bl}
\end{equation}

For $m < 0$, $\phi_{m}$ puts a black stone 
at position $-m$ (assuming a white stone is initially there), 
otherwise it annihilates the state 

\begin{equation}
\phi_{(m<0)}| j_1, j_2, \ldots \rangle = 
\left\{
\begin{array}{ll}
(-)^{k}|j_1, \ldots, j_{k}, -m, j_{k+1}, \ldots \rangle, & 
j_{k}<-m<j_{k+1} 
\bigskip  \\ 0, & \mbox{otherwise} 
\end{array}
\right.
\label{bm}
\end{equation}
and
\begin{equation}
\langle \ldots,i_2,i_1|\phi_{(m<0)} = 
\left\{
\begin{array}{ll}
(-)^{k-1} \langle \ldots, i_{k+1}, i_{k-1}, \ldots, i_1|, & m=i_k 
\bigskip \\ 0, & \mbox{otherwise} 
\end{array}
\right.
\label{bn}
\end{equation}

For $m=0$, $\phi_0$ acts on the site at the origin as follows

\begin{equation}
\phi_0|j_1,j_2,\ldots\rangle = \left\{
\begin{array}{ll}
\frac{1}{\sqrt{2}}|0, j_1, j_2 \ldots \rangle, & j_1 \not= 0 
\bigskip \\
\frac{1}{\sqrt{2}}| j_2, \ldots \rangle, & j_1 = 0
\end{array}
\right.
\end{equation}

\noindent and 

\begin{equation}
\langle \ldots, i_2, i_1| \phi_0 = \left\{
\begin{array}{ll}
\frac{1}{\sqrt{2}}\langle\ldots,i_2,i_1,0|, & i_1 \not= 0 \bigskip \\
\frac{1}{\sqrt{2}}\langle\ldots,i_2|, & i_1=0
\end{array}
\right.
\end{equation}

\subsection{The Lie algebra $B_\infty$ and neutral fermions} 
Following \cite{jimbo-miwa}, the Lie algebra ${B'}_\infty$, 
which is isomorphic to $B_\infty$, is generated by the bilinears
\begin{equation}
\left\{\sum_{i,j \in \Z}b_{ij}:\phi_{i}\phi_j:\right\}
\label{bo}
\end{equation}
\noindent where the coefficients $b_{ij}$ are constrained by 
the condition
\begin{equation*}
\exists\ N \in \N \ | \ b_{ij}=0, \ \forall \ |i+j|>N  
\end{equation*}

\subsection{$B_{\infty}$ Heisenberg subalgebra} Similarly to 
{\bf \ref{kp-1}}, we are interested in the Heisenberg subalgebra 
generated by $\lambda_m \in {B'}_\infty$, where

\begin{equation}
\lambda_m := \frac{1}{2}\sum_{j \in \Z}(-)^{j+1}
\phi_j\phi_{-j-m}, \quad m \in \Z_{\mbox{\tiny{odd}}}
\label{bp}
\end{equation}

\noindent satisfy the commutation relations

\begin{equation}
\left[ \lambda_m, \lambda_n \right] = \frac{m}{2}\delta_{m+n, 0}, 
\quad \forall\ m,n \in \Z_{\mbox{\tiny{odd}}}
\label{bq}
\end{equation}

\subsection{Two neutral fermion evolution operators} 
We begin by defining 

\begin{eqnarray*}
\Lambda_{\pm}(\mathbf{x}_{\mbox{\tiny{odd}}}) & := & 
\sum_{m \in \pm \N_{\mbox{\tiny{odd}}}} x_m \lambda_m \\ \\ \Phi(k) & := & \sum_{j \in \Z} \phi_j k^j
\end{eqnarray*}

Using 
$\left[ \lambda_m,\phi_n \right] = \phi_{n-m}$, 
$\forall\ m \in \Z_{\mbox{\tiny{odd}}}$,
$n \in \Z$,
it follows that 

\begin{equation*}
\left[\Lambda_{\pm}(\mathbf{x}_{\mbox{\tiny{odd}}}), 
\Phi(k) \right] = \sum_{m \in \pm \N_{\mbox{\tiny{odd}}}} 
x_m k^m \Phi(k) := \zeta_{\pm}(\mathbf{x}_{\mbox{\tiny{odd}}},k) 
\Phi(k)
\end{equation*}

The last commutator implies

\begin{equation}
e^{\Lambda_{\pm}(\mathbf{x}_{\mbox{\tiny{odd}}})}\Phi(k) 
e^{- \Lambda_{\pm}(\mathbf{x}_{\mbox{\tiny{odd}}})} =
\Phi(k) e^{\zeta_{\pm}(\mathbf{x}_{\mbox{\tiny{odd}}},k)}
\label{br}
\end{equation}

\subsection{Specializing the time variables}
Setting 

\begin{equation*}
x_m = \frac{2}{m}z^{-m}, \quad \forall\ m \in \Z_{\mbox{\tiny{odd}}}
\end{equation*}

\noindent and writing 
$\Lambda_{\pm}(\mathbf{x}_{\mbox{\tiny{odd}}}) := \Lambda_{\pm}(z)$,  
$\zeta_{\pm}(\mathbf{x}_{\mbox{\tiny{odd}}},k) := \zeta_{\pm}(z,k)$
under this specialization, 

\noindent we formally have

\begin{eqnarray}
\zeta_{+}(z,k) =
\sum_{m \in \N_{\mbox{\tiny{odd}}}}\frac{2}{m}
\ll \frac{k}{z} \rr^m 
&=&
\log{\ll \frac{z+k}{z-k} \rr} \label{bs} \\ \no \\ 
\zeta_{-}(z,k) = - \sum_{m \in \N_{\mbox{\tiny{odd}}}}\frac{2}{m}
\ll \frac{z}{k} \rr^m &=& \log{\ll \frac{k-z}{k+z} \rr} \no
\end{eqnarray}

\subsection{Neutral fermion vertex operators}
$\Gh^{+}(z)$ is defined as

\begin{equation}
\Gh_{+}(z) := e^{\Lambda_{+}(z)} =
\exp{\ll \sum_{ m \in \N_{\mbox{\tiny{odd}}}}
\frac{2}{m}z^{-m} \lambda_m \rr} 
\label{bt}
\end{equation}

\noindent and $\Gh_{-}(z)$ is defined as 

\begin{equation}
\Gh_{-}(z) := e^{-\Lambda_{-}(z)} =
\exp{\ll \sum_{ m \in \N_{\mbox{\tiny{odd}}}}
\frac{2}{m}z^m\lambda_{-m} \rr}  
\label{bu}
\end{equation}

Combining these definitions with the equations $(\ref{br})$ 
and $(\ref{bs})$, we have

\begin{eqnarray*}
\Gh_{+}(z) \Phi(k) \Gh_{+}(-z) & = & 
\Phi(k) \ll \frac{z+k}{z-k} \rr \\ \\ \\  
\Gh_{-}(-z) \Phi(k) \Gh_{-}(z) & = & 
\Phi(k) \ll \frac{k-z}{k+z} \rr
\end{eqnarray*}

These relations contain information about the time evolution of 
a neutral fermion (for specialized values of the time variables), 
which is revealed by writing the generating function $\Phi(k)$ in 
its sum form, and expanding formally:

\begin{eqnarray*}
\sum_{j \in \Z} \Gh_{+}(z) \phi_j 
\Gh_{+}(-z) k^j & = & 
\sum_{ j \in \Z} 
\phi_j k^j 
\ll
1 + 2 \sum_{n=1}^{\infty} 
\ll 
\frac{k}{z} 
\rr^n 
\rr \\ \\ \\
\sum_{j \in \Z} \Gh_{-}(-z) \phi_j 
\Gh_{-}(z) k^j & = & \sum_{ j \in \Z} 
\phi_j k^j 
\ll
1 + 2 \sum_{n=1}^{\infty}(-)^n 
\ll 
\frac{z}{k} 
\rr^n 
\rr
\end{eqnarray*}

Equating powers of $k$ in the previous expressions gives

\begin{eqnarray}
\Gh_{+}(z) \phi_j \Gh_{+}(-z) &=& 
\phi_j +2\sum_{n=1}^{\infty}\frac{1}{z^n}\phi_{j-n} \label{bv} \\ \no \\
\Gh_{-}(-z) \phi_j \Gh_{-}(z) &=& 
\phi_j +2\sum_{n=1}^{\infty} (-z)^n \phi_{j+n} 
\label{bw}
\end{eqnarray}

Given the definitions $(\ref{bt})-(\ref{bu})$ of the vertex operators

\begin{eqnarray*}
\Gh_{+}(z)
\Gh_{-}(z') &=& 
e^{\Lambda_{+}(z)}e^{-\Lambda_{-}(z')} \\ \\ 
&=&e^{[\Lambda_{+}(z),-\Lambda_{-}(z')]}\ e^{-\Lambda_{-}(z')}
e^{\Lambda_{+}(z)} \\ \\ &=&e^{[\Lambda_{+}(z),-\Lambda_{-}(z')]}\ 
\Gh_{-}(z')
\Gh_{+}(z)
\end{eqnarray*}

\noindent and given that

\begin{eqnarray*}
\left[ \Lambda_{+}(z), -\Lambda_{-}(z') \right] &=&
\sum_{m \in \N_{\mbox{\tiny{odd}}}}
\sum_{n \in \N_{\mbox{\tiny{odd}}}}
\frac{4}{mn}z^{-m}(z')^{n}
[
\lambda_m,\lambda_{-n}
] \\ \\ 
&=& \sum_{m \in \N_{\mbox{\tiny{odd}}}}
\sum_{n \in \N_{\mbox{\tiny{odd}}}}
\frac{4}{mn}z^{-m}(z')^{n}\frac{m}{2}\delta_{m,n} \\ \\ &=& 
\sum_{m \in \N_{\mbox{\tiny{odd}}}}
\frac{2}{m} \ll \frac{z'}{z}\rr^m = 
\log \ll \frac{z+z'}{z-z'} \rr
\end{eqnarray*}

\noindent we find

\begin{equation}
\Gh_{+}(z)
\Gh_{-}(z') = 
\ll
\frac{z+z'}{z-z'} 
\rr
\Gh_{-}(z')
\Gh_{+}(z)
\label{cd}
\end{equation}

\ni which is the basic commutation relation of neutral fermion
vertex operators.

\section{Strict Young diagrams}\label{bkp-2} 

Recalling the restriction on half-line Maya diagrams discussed 
in {\bf \ref{bkp-1}}, any half-line Maya diagram in the sector 
of initial states with the vacuum vector in Figure {\bf 
\ref{neutral-Maya-vacuum}} (which is the only sector that we are 
interested in), may be represented uniquely by

\begin{equation}
|\widehat{\mu}\r := \alpha(-)^{r}\phi_{m_1}\ldots 
\phi_{m_{2r}}|0\r = \alpha(-)^{r}
\dprod_{j=1}^{2r}\phi_{m_j} |0\r
\label{bx}
\end{equation}

\noindent whilst any half-line Maya diagram in the space of final 
states may be represented uniquely by

\begin{equation}
\l\widehat{\mu}| := \alpha(-)^{r+|\widehat{\mu}|}\l 0|\phi_{-m_{2r}}
\ldots\phi_{-m_1} = \alpha(-)^{r+|\widehat{\mu}|}\l 0|
\rprod_{j=1}^{2r}\phi_{-m_j}
\label{by}
\end{equation}

\noindent where $m_1 > \ldots > m_{2r} \geq 0$, $|\widehat{\mu}| = 
\sum_{j=1}^{2r}m_j$, and 

\begin{equation}
\alpha := \left\{
\begin{array}{ll}
1, & \quad m_{2r} \geq 1\\ \\ 
\sqrt{2}, & \quad m_{2r} = 0
\end{array}
\right.
\label{aaa}
\end{equation} 

The strict Young diagram $\widehat{\mu}=\{m_1,\ldots,m_{2r}\}$ 
corresponds to the initial state $(\ref{bx})$, and to the final 
state $(\ref{by})$, where the circumflex indicates that the Young 
diagram is strict: No two (non-zero length) parts have equal length. 

\subsection{Interlacing strict Young diagrams} One defines interlacing 
strict Young diagrams precisely as in the case of random (non-strict) 
Young diagrams.

\subsection{Generating interlacing strict Young diagrams}

\begin{lemma}
Let $|\widehat{\mu}\r$ and $\l\widehat{\mu} |$ be the initial state 
and final state representation, $(\ref{bx})$-$(\ref{by})$ respectively, 
of the strict partition $\widehat{\mu}=\{m_1,\ldots,m_{2r}\}$. Then 

\begin{equation}
\langle \widehat{\nu}|
\Gh_{+}(z)
| \widehat{\mu} \rangle 
= 
\left\{
\begin{array}{ll}
2^{n(\widehat{\nu}|\widehat{\mu})}
z^{|\widehat{\nu}|-|\widehat{\mu}|}, & 
\quad\widehat{\nu} \prec \widehat{\mu} \ 
\mbox{and}
\ n(\widehat{\nu})=n(\widehat{\mu}) 
\\ 
(-)^{n(\widehat{\mu})}
2^{n(\widehat{\nu}|\widehat{\mu})
+
\frac{1}{2}}z^{|\widehat{\nu}|-|\widehat{\mu}|}, 
&\quad 
\widehat{\nu} \prec \widehat{\mu} \ \mbox{and}\ n(\widehat{\nu})
=
n(\widehat{\mu})-1 
\\
0, & \quad \mbox{otherwise}
\end{array}
\right. 
\label{bz}
\end{equation}

\begin{equation}
\langle \widehat{\mu}|
\Gh{-}(z)
|\widehat{\nu}\rangle = 
\left\{
\begin{array}{ll}
2^{n(\widehat{\nu}|\widehat{\mu})}
z^{|\widehat{\mu}|-|\widehat{\nu}|}, & 
\quad
\widehat{\nu} \prec \widehat{\mu} \ \mbox{and}\ n(\widehat{\nu})=
n(\widehat{\mu}) 
\\ 
(-)^{n(\widehat{\mu})}2^{n(\widehat{\nu}|\widehat{\mu})+\frac{1}{2}}
z^{|\widehat{\mu}|-|\widehat{\nu}|}, &\quad \widehat{\nu} \prec 
\widehat{\mu} \ \mbox{and}\ n(\widehat{\nu})=n(\widehat{\mu})-1 
\\
0, & \quad \mbox{otherwise}
\end{array}
\right.
\label{ca}
\end{equation}

\noindent where $n(\widehat{\mu})$ denotes the number of non-zero 
elements in $\widehat{\mu}$, $n(\widehat{\nu}|\widehat{\mu})$ denotes 
the number of non-zero elements in $\widehat{\nu}$, not present in 
$\widehat{\mu}$.
\end{lemma}

\noindent{\bf Proof.} Set $m_{2r+1}\equiv-1$. Then we have

\begin{eqnarray}
\Gh_{+}(z)| \widehat{\mu} \r &=& \alpha (-)^r 
\Gh_{+}(z) 
 \dprod_{j=1}^{2r} \phi_{m_j} |0\r 
 \no
\\ 
\no
\\ 
&=& \alpha (-)^r \dprod_{j=1}^{2r} 
\ll
\Gh_{+}(z) \phi_{m_j}
\Gh_{+}(-z)
\rr
|0 \r
\no 
\\ 
\no
\\ 
&=& \alpha (-)^r 
\dprod_{j=1}^{2r}
\ll
\phi_{m_j}+2\sum_{i=1}^{\infty}
\frac{1}{z^i}
\phi_{(m_j-i)}
\rr
|0\r 
\label{something}
\end{eqnarray}
Consider the action of the product in the above equation, which we 
call $\widehat{P}$, on the vacuum $|0\r$. 
\begin{eqnarray*}
\widehat{P}|0\r
&=&
\ll
\phi_{m_1}+2\sum_{i=1}^{\infty}
\frac{1}{z^i}
\phi_{(m_1-i)}
\rr
\\
&\times&
\ll
\phi_{m_2}+2\sum_{i=1}^{\infty}
\frac{1}{z^i}
\phi_{(m_2-i)}
\rr
\ldots
\ll
\phi_{m_{2r}}+2\sum_{i=1}^{\infty}
\frac{1}{z^i}
\phi_{(m_{2r}-i)}
\rr
|0\r
\\
\end{eqnarray*}
Split the first sum from the left into parts and rewrite it as
\begin{eqnarray*}
\widehat{P}|0\r
&=&
\ll
\phi_{m_1}+2\sum_{i=1}^{m_1-m_2-1}
\frac{1}{z^i}
\phi_{(m_1-i)}
+
\frac{1}{z^{m_1-m_2}}
\phi_{m_2}
\right.
\\
&+&
\left.
\frac{1}{z^{m_1-m_2}}
(
\phi_{m_2}
+
2\sum_{i=1}^{\infty}
\frac{1}{z^i}\phi_{(m_2-i)}
)
\rr
\prod_{j=2}^{2r}
\ll
\phi_{m_{j}}+2\sum_{i=1}^{\infty}
\frac{1}{z^i}
\phi_{(m_{j}-i)}
\rr
|0\r
\end{eqnarray*}
Using the identity
\begin{equation*}
\ll
\phi_{m}+2\sum_{i=1}^{\infty}
\frac{1}{z^i}
\phi_{(m-i)}
\rr
\ll
\phi_{m}+2\sum_{i=1}^{\infty}
\frac{1}{z^i}
\phi_{(m-i)}
\rr
=
0
\end{equation*}
which follows by expanding each sum (including all terms up to 
$\phi_{-m}$ as all of these will contribute to the required
result on the left hand side) and using the anti-commutation 
relation (\ref{neutralanti}), we obtain
\begin{eqnarray*}
\widehat{P}|0\r
&=&
\ll
\phi_{m_1}+2\sum_{i=1}^{m_1-m_2-1}
\frac{1}{z^i}
\phi_{(m_1-i)}
+
\frac{1}{z^{m_1-m_2}}
\phi_{m_2}
\rr
\\
& \times &
\ll
\phi_{m_2}+2\sum_{i=1}^{\infty}
\frac{1}{z^i}
\phi_{(m_2-i)}
\rr
\ldots
\ll
\phi_{m_{2r}}+2\sum_{i=1}^{\infty}
\frac{1}{z^i}
\phi_{(m_{2r}-i)}
\rr
|0\r
\end{eqnarray*}
This procedure can then be performed on the second sum from the left, 
and so on, until one reaches the last sum, which truncates because 
$\phi_m$ annihilates $|0\r$ for all $m < 0$. Hence
\begin{equation*}
\widehat{P}|0\r
=
\dprod_{j=1}^{2r}
\ll
\phi_{m_j} + 2 \sum_{i=1}^{m_j-m_{j+1}-1}
\frac{1}{z^i} \phi_{(m_j-i)} + \frac{1}{z^{m_j-m_{j+1}}}
\phi_{m_{j+1}}
\rr
|0\r 
\end{equation*}
Using the above result in (\ref{something}), we obtain

\begin{eqnarray}
\Gh_{+}(z)| \widehat{\mu} \r &=&
\alpha (-)^r 
\dprod_{j=1}^{2r}
\ll
\phi_{m_j} + 2 \sum_{i=1}^{m_j-m_{j+1}-1}
\frac{1}{z^i} \phi_{(m_j-i)} + \frac{1}{z^{m_j-m_{j+1}}}
\phi_{m_{j+1}}
\rr
|0\r 
   \nonumber
\\ \nonumber
\\ 
&=& 
\sum_{\substack{\widehat{\nu} \prec \widehat{\mu} 
\\  
\\  
n(\widehat{\nu}) = n(\widehat{\mu})}} 
2^{n(\widehat{\nu}| \widehat{\mu})}z^{|\widehat{\nu}|
-|\widehat{\mu}|}
|\widehat{\nu}\rangle+(-)^{n(\widehat{\mu})}\sqrt{2}
\sum_{\substack{\widehat{\nu} \prec \widehat{\mu} 
\\  
\\  
n(\widehat{\nu})=n(\widehat{\mu})-1}}
2^{n(\widehat{\nu}|\widehat{\mu})}
z^{|\widehat{\nu}|-|\widehat{\mu}|}|
\widehat{\nu}\rangle 
\label{result2}
\end{eqnarray}

It is instructive to figure out the origin of the factors of 2 in 
(\ref{result2}). Recall that $\widehat{\mu}$ is a strict partition 
with parts of lengths $m_1 > \cdots > m_{2r} \geq 0$, and that 
the action of $\Gh_{+}(z)$ reduces the heights of the parts of 
$\widehat{\mu}$ or leaves them invariant. Consider the right hand 
side of first line of (\ref{result2}). The $j$-th factor in the 
product describes the evolution of the $j$-th part of $\widehat{\mu}$, 
which has length $m_j$. From the forms of the terms in this factor, 
we see that there are 3 possibilities: 

The first term leads to a part of length $m_j$, so there is no 
change in length, and no new factors are introduced. 
The second term leads to a part of length $m_{j} - i$, 
$1 \leq i \leq (m_j - m_{j+1} - 1)$ which does not occur in 
$\widehat{\mu}$, a factor of $z^{-i}$ is introduced (that keeps 
track of the change in length) as well as a factor of 2. 
The third term leads to a part of length $m_{j+1}$, which is 
different from the original length of $m_{j}$ but is a length 
that occurs in $\widehat{\mu}$, a factor of $z^{-m_j + m_{j+1}}$
is introduced, but no factor of 2. 

What we learn from this is that, while powers of $z$ keep track 
of changes in lengths, every time that a totally new length appears, 
we acquire an extra factor of 2. This explains the factors of 2 in 
the second line of (\ref{result2}).

Further, notice that the result of (\ref{result2}) has two terms.
The first is a sum over all partitions $\widehat{\nu}$ that have 
the same number of non-zero parts as $\widehat{\mu}$. The second 
is a sum over all partitions $\widehat{\nu}$ with one non-zero 
part less than $\widehat{\mu}$. 

The factor of $(-)^{n(\widehat{\mu})} \sqrt{2}$ in the second term 
accounts for the relative minus sign and valuation of $\alpha$, 
as defined in (\ref{aaa}), between the two types of partitions. 

Analogously to the proof of (\ref{result2}), we have 

\begin{eqnarray*}
\l 
\widehat{\mu}|\Gh_{-}(z) 
&=& 
\alpha 
(-)^{r+|\widehat{\mu}|}\l 0|
\rprod_{j=1}^{2r}\phi_{-m_j}
\Gh_{-}(z) 
= \alpha (-)^{r+|\widehat{\mu}|}\l 0|
\rprod_{j=1}^{2r}
\ll
\Gh_{-}(-z)\phi_{-m_j}\Gh_{-}(z)
\rr
\\ 
\\ 
&=& \alpha (-)^{r+|\widehat{\mu}|}\l 0|\rprod_{j=1}^{2r}
\ll
\phi_{-m_j}+2\sum_{i=1}^{\infty}(-z)^i \phi_{(-m_j+i)}
\rr
\\ 
\\ 
&=& 
\alpha (-)^{r+|\widehat{\mu}|}\l 0|\rprod_{j=1}^{2r}
\ll
\phi_{-m_j  } + 2 \sum_{i=1}^{m_j-m_{j+1}-1}(-z)^i 
\phi_{(-m_j+i)} + (-z)^{m_j-m_{j+1}}\phi_{-m_{j+1}}
\rr
\\ 
\\ 
&=& 
\sum_{\substack{\widehat{\nu} \prec \widehat{\mu} 
\\ 
\\ 
n(\widehat{\nu})=n(\widehat{\mu})}}
2^{n(\widehat{\nu}|\widehat{\mu})}
z^{|\widehat{\mu}|-|\widehat{\nu}|}
\l\widehat{\nu}| + (-)^{n(\widehat{\mu})}
\sqrt{2}
\sum_{\substack{\widehat{\nu} \prec \widehat{\mu} 
\\ 
\\ 
n(\widehat{\nu})=n(\widehat{\mu})-1}}
2^{n(\widehat{\nu}|\widehat{\mu})}
z^{|\widehat{\mu}|-|\widehat{\nu}|}
\l
\widehat{\nu}| 
\quad\quad\proofend
\end{eqnarray*}

The proof is completed by taking appropriate inner products.

\section{Diagonally strict plane partitions}\label{bkp-3}

\begin{de} A diagonally strict plane partition $\widehat{\pi}$ 
is a plane partition whose diagonal slices, given by

\begin{equation*}
\widehat{\mu}_m = \left\{
\begin{array}{ll}
\{ \widehat{\pi}(-m+1,1),\ 
\widehat{\pi}(-m+2,2),\ldots\}, & \quad m \leq 0 \\ \\
\{ \widehat{\pi}(1,m+1),\ 
\widehat{\pi}(2,m+2),\ldots\}, & \quad m \geq 0
\end{array}
\right.
\end{equation*}

\noindent are all strict Young diagrams.
\end{de}

An example of a diagonally strict plane partition is in Figure 
{\bf \ref{strict-plane-partition}}. The diagonals are shown on a planar 
representation in Figure {\bf \ref{strict-slices}}.

%
\begin{center}
\begin{minipage}{4.9in}
\setlength{\unitlength}{0.0012cm}
\renewcommand{\dashlinestretch}{30}
\begin{picture}(4800, 5000)(-2000, 0)
\thicklines
\path(3912,1512)(3312,1512)
\path(1812,4212)(2412,4212)
\path(1812,4212)(1512,3912)
\path(2412,4212)(2112,3912)
\path(1512,3912)(2112,3912)
\path(1512,3912)(1512,3312)
\path(2112,3912)(2112,3312)
\path(1512,3312)(2112,3312)
\path(2412,4212)(2412,3612)
\path(2412,3612)(2112,3312)
\path(1512,3312)(1212,3012)
\path(2112,3312)(1812,3012)
\path(1212,3012)(1812,3012)
\path(1212,3012)(1212,2412)
\path(1812,3012)(1812,2412)
\path(1212,2412)(1812,2412)
\path(1212,2412)(0912,2112)
\path(1812,2412)(1512,2112)
\path(0912,2112)(1512,2112)
\path(0912,2112)(0912,1512)
\path(1512,2112)(1512,1512)
\path(0912,1512)(1512,1512)
\path(0912,1512)(0612,1212)
\path(1512,1512)(1212,1212)
\path(0612,1212)(1212,1212)
\path(0612,1212)(0612,0612)
\path(1212,1212)(1212,0612)
\path(0612,0612)(1212,0612)
\path(0612,0612)(0012,0012)
\path(2112,3312)(2112,2712)
\path(2112,2712)(1812,2412)
\path(1812,2412)(1812,1812)
\path(1812,1812)(1512,1512)
\path(1512,1512)(1512,0912)
\path(1512,0912)(1212,0612)
\path(2412,3612)(2412,3012)
\path(2412,3012)(2112,2712)
\path(2412,3012)(3012,3012)
\path(2112,2712)(2712,2712)
\path(3012,3012)(2712,2712)
\path(1812,2412)(2412,2412)
\path(2712,2712)(2412,2412)
\path(2412,2412)(2412,1812)
\path(1812,1812)(2412,1812)
\path(2412,1812)(2112,1512)
\path(1512,1512)(2112,1512)
\path(2112,1512)(2112,0912)
\path(1512,0912)(2112,0912)
\path(3012,3012)(3012,2412)
\path(2712,2712)(2712,2112)
\path(3012,2412)(2412,1812)
\path(3012,2412)(4512,2412)
\path(3612,2412)(2712,1512)
\path(2712,2112)(3312,2112)
\path(2412,1812)(3012,1812)
\path(2112,1512)(2712,1512)
\path(2712,1512)(2712,0912)
\path(2112,0912)(2712,0912)
\path(4212,2412)(3912,2112)
\path(4512,2412)(4812,2412)
\path(4812,2412)(4512,2112)
\path(3312,2112)(4512,2112)
\path(3012,1812)(3012,1212)
\path(3912,2112)(3912,1512)
\path(4512,2112)(4512,1512)
\path(4812,2412)(4812,1812)
\path(4812,1812)(4512,1512)
\path(4812,1812)(5412,1812)
\path(3012,1212)(2712,0912)
\path(3912,1512)(4512,1512)
\path(3312,2112)(3312,1512)
\path(3312,1512)(3012,1212)
\path(1812,4812)(1812,4212)
\end{picture}
\begin{ca}
\label{strict-plane-partition}
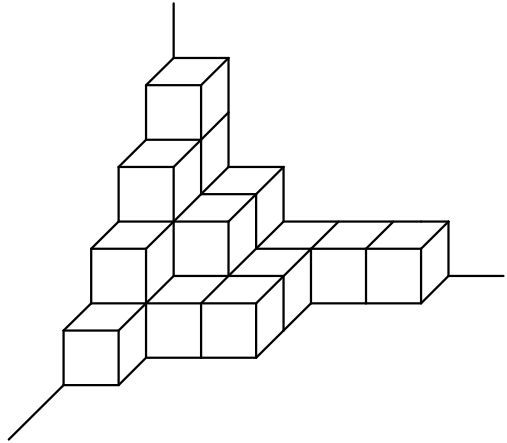
A diagonally strict plane partition. Notice that all connected 
horizontal plateaux are at most 1 square wide: there is at most 
one way to move from one square to another square at the same
level, without changing levels.
\end{ca}
\end{minipage}
\end{center}
\bigskip

%
\begin{center}
\begin{minipage}{4.9in}
\setlength{\unitlength}{0.0012cm}
\renewcommand{\dashlinestretch}{30}
\begin{picture}(4800, 3600)(-2000, 0)
\thicklines
\path(0000,0600)(0600,0600)
\path(0000,1200)(1800,1200)
\path(0000,1800)(1800,1800)
\path(0000,2400)(3000,2400)
\path(0000,3000)(3000,3000)
\path(0000,3000)(0000,0600)
\path(0600,3000)(0600,0600)
\path(1200,3000)(1200,1200)
\path(1800,3000)(1800,1200)
\path(2400,3000)(2400,2400)
\path(3000,3000)(3000,2400)
\thinlines
\path(0000,1200)(0900,0300)
\path(0000,1800)(0900,0900)
\path(0000,2400)(1500,0900)
\path(0000,3000)(2100,0900)
\path(0600,3000)(2100,1500)
\path(1200,3000)(2100,2100)
\path(1800,3000)(2700,2100)
\path(2400,3000)(3300,2100)
\put(0300,0900){1}
\put(0300,1500){2}
\put(0300,2100){3}
\put(0300,2700){4}
\put(0900,1500){1}
\put(0900,2100){2}
\put(0900,2700){2}
\put(1500,1500){1}
\put(1500,2100){1}
\put(1500,2700){1}
\put(2100,2700){1}
\put(2700,2700){1}
\put(0900,0000){$\widehat{\mu}_{-3}$}
\put(0900,0600){$\widehat{\mu}_{-2}$}
\put(1500,0600){$\widehat{\mu}_{-1}$}
\put(2100,0600){$\widehat{\mu}_{ 0}$}
\put(2100,1200){$\widehat{\mu}_{ 1}$}
\put(2100,1800){$\widehat{\mu}_{ 2}$}
\put(2700,1800){$\widehat{\mu}_{ 3}$}
\put(3300,1800){$\widehat{\mu}_{ 4}$}
\end{picture}
\begin{ca}
\label{strict-slices}
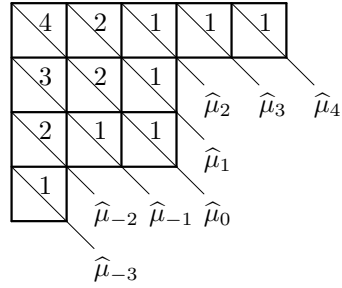
A planar representation of the diagonally strict plane partition in 
Figure {\bf \ref{strict-plane-partition}} and its diagonal slices. 
The integers in the boxes are the heights of the corresponding columns. 
\end{ca}
\end{minipage}
\end{center}
\bigskip

\subsection{Strict diagonal slices}

\begin{de} Given a diagonally strict plane partition $\widehat{\pi}$, 
consider only those points $(i,j)$ for which $\widehat{\pi}(i,j)\not=0$. 
We say that a path begins at the coordinate pair $(i,j)$ if the 
conditions
$\widehat{\pi} (i+1, j) \not= \widehat{\pi} (i, j)$, and 
$\widehat{\pi} (i, j-1) \not= \widehat{\pi} (i, j)$
are both satisfied, where 
$\widehat{\pi} (i,0) \equiv 0, \ \forall\ i$. 
A coordinate pair $(i,j)$ that does not satisfy the preceding criteria is 
a member of a pre-existing path. We shall henceforth let $p(\widehat{\pi})$ 
denote the number of paths possessed by $\widehat{\pi}$.
\end{de}

\noindent Consider the scalar product

\begin{eqnarray}
S_{B}(q) &:=& \langle 0|
{\rprod_{j=1}^{\infty}
\Gh_{+}\ll q^{\frac{-2j+1}{2}}\rr}\ 
{\dprod_{k=1}^{\infty}
\Gh_{-}\ll q^{\frac{2k-1}{2}}\rr}|0
\rangle \label{cb} 
\\ 
\no \\ 
&=&
\sum_{\widehat{\mu}}\langle 0|{\rprod_{j=1}^{\infty}
\Gh_{+}\ll q^{\frac{-2j+1}{2}}\rr}|\widehat{\mu}\rangle\langle
\widehat{\mu}|\ {\dprod_{k=1}^{\infty}
\Gh_{-}\ll q^{\frac{2k-1}{2}}\rr}|0\rangle \label{cc}
\end{eqnarray}

\noindent where $q$ is an indeterminate and $\sum_{\widehat{\mu}}$ 
denotes a sum over all strict Young diagrams $\widehat{\mu}$. We have 
seen that $\Gh_{+}(z)|\widehat{\mu}\rangle$ and 
$\langle\widehat{\mu}|\Gh_{-}(z)$ generate all strict 
Young diagrams $\widehat{\nu} \prec \widehat{\mu}$, with weightings 
given by Lemma {\bf 2}. It follows that equation $(\ref{cc})$ 
generates a weighting equal to 

\begin{equation*}
{\rprod_{j=1}^{M}\langle\widehat{\nu}_{-j}|
\Gh_{+}\ll q^{\frac{-2j+1}{2}}\rr}|
\widehat{\nu}_{-j+1}\rangle\ 
{\dprod_{k=1}^{N}\langle\widehat{\nu}_{k-1}|
\Gh_{-}\ll q^{\frac{2k-1}{2}}\rr}|
\widehat{\nu}_k\rangle = 2^{p(\widehat{\pi})}
\prod_{j=-M}^{N}q^{|\widehat{\nu}_j|}
\end{equation*}

\noindent for all diagonally strict plane partitions given by

\begin{equation*}
\widehat{\pi}=\left\{\emptyset=\widehat{\nu}_{-M} \prec \ldots \prec 
\widehat{\nu}_{-1} \prec \widehat{\nu}_0 \succ \widehat{\nu}_1 \succ \ldots 
\succ \widehat{\nu}_N=\emptyset \right\} 
\end{equation*}

\noindent
Notice that, as explained earlier, due to the powers of 2 appearing in 
Lemma {\bf 2}, starting at the main diagonal slice of $\widehat{\pi}$ 
and working outwards, a factor of 2 is acquired for every path in 
$\widehat{\pi}$. This explains the weighting of $2^{p(\widehat{\pi})}$ 
in the above equation. Hence

\begin{equation*}
S_{B}(q)=\sum_{\widehat{\pi}}2^{p(\widehat{\pi})}q^{|\widehat{\pi}|}
\end{equation*}  

\noindent Applying the commutation relation $(\ref{cd})$ successively to 
equation $(\ref{cb})$, one recovers a product form for the generating 
function $S_{B}(q)$:

\bigskip 
\begin{boxedminipage}[c]{12cm}

\begin{equation}
S_{B}(q) = \prod_{n=1}^{\infty}
\ll
\frac{1+q^n}{1-q^n}
\rr^n
\label{ce}
\end{equation}

\end{boxedminipage}
\bigskip 

\subsection{The neutral fermion two-dimensional Toda lattice hierarchy}
The generating function (\ref{ce}) is a specialization of 
\begin{equation}
S_B({\bf x}, {\bf y}) = \sum_{\widehat{Y}} 
2^{-n(\widehat{Y})}
Q_{\widehat{Y}}(x_1,x_2,\ldots)
Q_{\widehat{Y}}(y_1,y_2,\ldots)
=
\prod_{i,j=1}^{\infty}
\frac{
1+x_i y_j
}
{
1-x_i y_j
}
\end{equation}
\noindent where $Q_{\widehat{Y}}$ is the Schur $Q$-function associated 
to the strict Young diagram $\widehat{Y}$ \cite{macdonald}, and 
$n(\widehat{Y})$ is the number of non-zero elements in $\widehat{Y}$, 
to 
$\{{\bf x}\}$
$=$
$ \{{\bf y}\}$
$=$ 
$\{ q^{\frac{1}{2}}, $$q^{\frac{3}{2}}, $$\ldots \}$. We postulate that 
under a suitable change of variables, $S_B$ is a tau function of the 
neutral fermion analogue of the Toda hierarchy of the type studied in 
\cite{Or,OrNi}. 

\section*{Acknowledgements}
OF wishes to thank Prof T~Shiota and Prof K~Takasaki for discussions. 
MW and MZ are supported by Australian Postgraduate Awards.


\end{document}